\documentclass[11pt]{article}
\usepackage{amsfonts,amsmath,amssymb}
\usepackage{enumerate}
\usepackage{hyperref}
\usepackage{bbm}
\usepackage{nicefrac}
\usepackage[all]{xy}
\usepackage{graphicx}
\usepackage{bm}
\usepackage{makecell}
\usepackage[table]{xcolor}
\usepackage{longtable}		
\usepackage{subcaption}		
\usepackage{tikz}			

\usepackage{upgreek}

\usepackage{float}			
\usepackage{lscape}                 

\usepackage{booktabs}

\def\sst#1{{\scriptscriptstyle #1}}

\def\0{{\sst{(0)}}}
\def\1{{\sst{(1)}}}
\def\2{{\sst{(2)}}}
\def\3{{\sst{(3)}}}
\def\4{{\sst{(4)}}}
\def\5{{\sst{(5)}}}
\def\6{{\sst{(6)}}}
\def\7{{\sst{(7)}}}
\def\8{{\sst{(8)}}}

\allowdisplaybreaks  

\usepackage{multirow}
\usepackage{rotating}

\allowdisplaybreaks

\newcommand{\cN}{\mathcal{N}}

\usepackage{lipsum}

\newlength\Colsep
\begin{document}

\makeatletter
\renewcommand{\theequation}{\thesection.\arabic{equation}}
\@addtoreset{equation}{section}
\makeatother

\begin{titlepage}

\begin{flushright}
IFT-UAM/CSIC-22-60 \\
%
\end{flushright}

\vspace{5pt}

   \begin{center}
   \baselineskip=16pt

   \begin{Large}\textbf{
\hspace{-18pt} $\mathcal{N}=1$ S-fold spectroscopy   \\[8pt]
}
   \end{Large}

\vspace{25pt}

{\large  Mattia Ces\`aro$^{1}$, Gabriel Larios$^{1}$ \,and \,  Oscar Varela$^{1,2}$}
		
\vspace{25pt}

	\begin{small}

	{\it $^{1}$ Departamento de F\'\i sica Te\'orica and Instituto de F\'\i sica Te\'orica UAM/CSIC , \\
   Universidad Aut\'onoma de Madrid, Cantoblanco, 28049 Madrid, Spain}  \\

	\vspace{15pt}
	
	{\it $^{2}$ Department of Physics, Utah State University, Logan, UT 84322, USA}     \\	
		
	\end{small}

\vskip 50pt

\end{center}

\begin{center}
\textbf{Abstract}
\end{center}

\begin{quote}

We analyse the spectrum of Kaluza-Klein excitations above three distinct families of ${\cal N}=1$ AdS$_4$ solutions of type IIB supergravity of typically non-geometric, S-fold type that have been recently found. For all three families, we provide the complete algebraic structure of their spectra, including the content of OSp$(4|1)$ multiplets at all Kaluza-Klein levels and their charges under the residual symmetry groups. We also provide extensive results for the multiplet dimensions using new methods derived from exceptional field theory, including complete, analytic results for one of the families. All three spectra show periodicity in the moduli that label the corresponding family of solutions. Finally, the compactness of these moduli is verified in some cases at the level of the fully-fledged type IIB uplifted solutions.

\end{quote}

\vfill

\end{titlepage}

\tableofcontents



\section{Introduction}


New anti-de Sitter (AdS) vacua of maximal supergravity in four dimensions with the specific gauge groups 
\begin{equation} \label{eq:GaugeGroups}
G_- \equiv [{\rm SO(6)}\times{\rm SO(1,1)}]\ltimes\mathbb{R}^{12} 
\qquad
\textrm{or} \qquad G_+ \equiv [{\rm SO(6)}\times{\rm SO(2)}]\ltimes\mathbb{R}^{12} \; ,
\end{equation}
have been recently found \cite{Gallerati:2014xra,Guarino:2019oct,Guarino:2020gfe,Bobev:2021yya,Guarino:2021hrc,Berman:2021ynm,Bobev:2021rtg}. These vacua are interesting because they are in one-to-one correspondence with type IIB supergravity solutions of the form 
\begin{equation} \label{eq:AdS4ixS5xS1}
\textrm{AdS}_4 \times S^5 \times S^1 \; ,
\end{equation}
via consistent uplift \cite{Inverso:2016eet} (see also \cite{Berman:2021ynm}). On this type of solutions, the type IIB fields that lie in non-trivial representations of the SL$(2, \mathbb{R})$ S-duality group often take on non-trivial monodromies around the $S^1$ factor in (\ref{eq:AdS4ixS5xS1}) \cite{Inverso:2016eet}. For this reason, the resulting solutions are typically non-geometric and go under the name of S-folds. For certain solutions (\ref{eq:AdS4ixS5xS1}) that uplift from the $G_+$ gauging, the monodromies are trivial, though, and the solutions are perfectly geometric \cite{Berman:2021ynm}. These type IIB configurations generically are limiting cases of Janus-type solutions \cite{DHoker:2007zhm} and, therefore, the dual three-dimensional conformal field theories (CFTs) should arise as limits of four-dimensional $\cN=4$ super-Yang-Mills at codimension-one defects \cite{Gaiotto:2008sd}. See \cite{Bobev:2019jbi,Bobev:2020fon,Arav:2021tpk,Arav:2021gra} for alternative supergravity constructions that also lead to this type of S-folds.

A characteristic feature of the vacua of the $D=4$ $\cN=8$ gaugings (\ref{eq:GaugeGroups}) is that they tend to arise as critical {\it loci} on the $\textrm{E}_{7(7)} / \textrm{SU}(8)$ scalar manifold. This is unlike other gaugings \cite{deWit:1982ig,Guarino:2015qaa} with similar higher-dimensional origin \cite{deWit:1986iy,Guarino:2015jca,Guarino:2015vca}, whose vacua correspond instead to isolated critical {\it points} of the corresponding scalar potentials. In other words, the vacua of the $G_\pm$ gaugings (\ref{eq:GaugeGroups}), all known of which are AdS, come in families parameterised by (pseudo)scalars of the $D=4$ $\cN=8$ supergravity that remain massless. Thus, the corresponding families of S-folds (\ref{eq:AdS4ixS5xS1}) should be holographically dual to (subsectors of) conformal manifolds of the boundary CFTs. This feature makes of the gaugings (\ref{eq:GaugeGroups}) excellent starting points to study holographically this class of conformal manifolds. In contrast, for solutions that uplift from isolated critical points \cite{deWit:1986iy,Guarino:2015jca,Guarino:2015vca}, specific methods \cite{Lunin:2005jy,Ashmore:2016oug} beyond gauged supergravity need to be employed in order to study their conformal manifolds, see {\it e.g.}~\cite{Bobev:2021gza,Ashmore:2021mao}. 

For example, the $G_-$ gauging has an AdS critical point with residual supersymmetry $\cN =4$ and bosonic symmetry SO(4) \cite{Gallerati:2014xra}. This gives rise to a type IIB S-fold (\ref{eq:AdS4ixS5xS1}) with the same (super)symmetries \cite{Inverso:2016eet}, whose dual CFT was identified in \cite{Assel:2018vtq}. The $\cN=4$ SO(4) critical point \cite{Gallerati:2014xra} is not isolated, though: it belongs to a two-parameter family of AdS solutions with generic supersymmetry $\cN =2$ and $\textrm{U}(1)^2 \subset \textrm{SO}(4)$ bosonic symmetry \cite{Bobev:2021yya} (see also \cite{Guarino:2020gfe,Arav:2021gra} for previously constructed subsectors of this $\cN=2$  family and \cite{Guarino:2021hrc} for a non-supersymmetric extension). The two parameters that characterise the family of solutions are (pseudo)scalars of the $D=4$ $\cN =8$ supergravity that remain massless. Thus, although the type IIB uplift (\ref{eq:AdS4ixS5xS1}) of the entire two-parameter family has not been constructed in full generality (see \cite{Inverso:2016eet,Guarino:2020gfe,Giambrone:2021zvp,Cesaro:2021tna} for partial results), it is guaranteed to exist by consistent uplift \cite{Inverso:2016eet}. The resulting two-parameter family of type IIB S-folds should thus be dual to the conformal manifold of the $\cN=4$ CFT of \cite{Assel:2018vtq}. 

In this paper, we turn our attention to the remaining known supersymmetric families of AdS vacua of the $D=4$ $\cN=8$ gaugings (\ref{eq:GaugeGroups}). These include three instances, all of them strictly with $\cN=1$ supersymmetry: one in the $G_-$ gauging \cite{Guarino:2019oct,Guarino:2020gfe} and two in the $G_+$ gauging \cite{Berman:2021ynm,Bobev:2021rtg}. See section \ref{sec:AdSSols} for a review. The family of AdS solutions in the $G_-$ gauging is two-parametric and generically U$(1)^2$-invariant. It displays symmetry enhancements to $\textrm{SU}(2) \times \textrm{U}(1)$ at one-dimensional subloci and to SU(3) at a point. Both families in the $G_+$ gauging are one-parametric and generically U(1) invariant. One of them presents a symmetry enhancement to SO(3) at a point. The type IIB uplifts (\ref{eq:AdS4ixS5xS1}) of the U$(1)^2$-invariant family \cite{Guarino:2021kyp,Guarino:2019oct} and of the SO(3) point \cite{Berman:2021ynm} are known. These solutions should be dual to infrared conformal fixed surfaces, lines and points of four-dimensional $\cN=4$ super-Yang-Mills theory at codimension-one defects \cite{Guarino:2021kyp,Arav:2021gra}.

Specifically, we compute the spectrum of Kaluza-Klein (KK) perturbations above these families of $\cN=1$ AdS$_4$ solutions, thereby extending to these cases the recent KK analysis \cite{Cesaro:2021tna,Giambrone:2021zvp} for the $\cN=2$ holographic conformal manifold \cite{Bobev:2021yya} of the $\cN=4$ point \cite{Gallerati:2014xra,Inverso:2016eet}. Our results assess holographically the spectrum of single trace operators on the marginally deformed $\cN=1$ dual CFTs. For all three families of solutions, we characterise their spectra by listing their complete algebraic structure in terms of $\textrm{OSp}(4|1) \times G$ representations at all KK levels. Here, $G$ stands for any of the residual symmetry groups, U$(1)^2$, U(1), etc., mentioned above. Except for a few massless cases that we identify, all the $\textrm{OSp}(4|1)$ multiplets in the spectra are long. For this reason, the dimensions of (the superconformal primaries in) these multiplets are not fixed by the above algebraic structure and need to be computed independently. In order to do this, we have diagonalised the bosonic \cite{Malek:2019eaz,Malek:2020yue} (see also \cite{Varela:2020wty}) and fermionic  \cite{Cesaro:2020soq} mass matrices recently derived for the KK perturbations on the AdS solutions of the higher-dimensional supergravities that uplift consistently \cite{deWit:1982ig,Guarino:2015qaa,Inverso:2016eet,Berman:2021ynm} from $D=4$ $\cN=8$ gauged supergravities. These KK mass matrices were obtained in those references with the help of exceptional field theory (ExFT) \cite{Hohm:2013pua,Hohm:2013uia,Godazgar:2014nqa}, a duality-covariant reformulation of the higher-dimensional supergravities (see \cite{Berman:2020tqn} for a review). 

Section \ref{sec:spectra} contains the main results of our KK analysis, and further details can be found in the appendices. Remarkably, for the U$(1)^2$-invariant family we are able to give closed formulae for the supermultiplet dimensions at all KK levels. For the other two families we also provide some analytic results for the dimensions, but most of them are numerical: we give the multiplet dimensions for the first few KK levels at discretised values of the moduli that label each family. In all cases, the dependence of the supermultiplet dimensions on the moduli is locked into fixed combinations that involve one of the two KK levels and the $G$ charges. Together with the independence on this KK level of the algebraic structure of the spectra, this fact is responsible for the periodicity of the latter in moduli space. This compact character of the moduli is intrinsically ten-dimensional and cannot be seen at the level of the $D=4$ $\cN=8$ supergravity. This compactness has been previously observed for one \cite{Giambrone:2021zvp}, but interestingly not the other \cite{Cesaro:2021tna}, of the two moduli of the holographic conformal manifold \cite{Bobev:2021yya} of the $\cN=4$ S-fold \cite{Gallerati:2014xra,Inverso:2016eet} in this class.

This paper extends previous partial results on the KK spectrum of the $\cN=1$ AdS solutions at hand, including the lowest-lying spectra \cite{Guarino:2020gfe,Berman:2021ynm,Bobev:2021rtg} and the spin-two spectrum \cite{Dimmitt:2019qla} at the SU(3) point. Together with \cite{Giambrone:2021zvp,Cesaro:2021tna}, we exhaust the KK spectra of all known (to date) supersymmetric type IIB S-fold solutions of the form (\ref{eq:AdS4ixS5xS1}) that uplift from $D=4$ $\cN=8$ supergravity with gaugings (\ref{eq:GaugeGroups}). Together with \cite{Giambrone:2021zvp,Cesaro:2021tna} and 
\cite{Englert:1983rn,Sezgin:1983ik,Biran:1983iy,Klebanov:2008vq,Malek:2020yue,Varela:2020wty,Cesaro:2020soq,Cesaro:2021haf}, we also exhaust the KK spectra of all known supersymmetric AdS$_4$ solutions of M-theory or type II that uplift consistently \cite{deWit:1986iy,Guarino:2015jca,Guarino:2015vca,Inverso:2016eet} from $D=4$ $\cN=8$ gauged supergravities. Other KK spectra of related AdS solutions that have been recently computed using the ExFT-derived methods of \cite{Malek:2019eaz,Cesaro:2020soq} include \cite{Malek:2020mlk,Guarino:2020flh,Eloy:2020uix,Bobev:2020lsk,Eloy:2021fhc,Giambrone:2021wsm}.


\section{Solutions} \label{sec:AdSSols}


We start by reviewing the $\cN=1$ AdS vacua of the gauged supergravities under consideration \cite{Guarino:2019oct,Guarino:2020gfe,Berman:2021ynm,Bobev:2021rtg} in order to fix our notation.

\subsection{Common aspects}

We are interested in a specific class of type IIB S-folds that uplift from vacua of two concrete gaugings of $D=4$ $\cN=8$ supergravity. These have dyonically-gauged, in the sense of \cite{DallAgata:2011aa,Dall'Agata:2014ita,Inverso:2015viq}, gauge groups (\ref{eq:GaugeGroups}), and are characterised by an embedding tensor \cite{deWit:2007mt} (see \cite{Trigiante:2016mnt} for a review) in the $\bm{36}$ and $\bm{36'}$ of $\textrm{SL}(8, \mathbb{R})$,
\begin{equation}	\label{eq: embtensors}
	\Theta_{AB}{}^C{}_D=2\delta_{[A}^C\theta_{B]D}\,,	\qquad\qquad
	\Theta^{ABC}{}_D=2\delta^{[A}_D\xi^{B]C}\, .
\end{equation}
Here, $A=1,\dots,8$ is a fundamental index of ${\rm SL}(8,\mathbb{R})$ and
\begin{equation}	\label{eq: embtensorcomponents}
	\theta \equiv g \,  {\rm diag}\big(0,1,1,1,1,1,1,0\big)\,,	\qquad\qquad
	\xi \equiv m \,   {\rm diag}\big(x,0,0,0,0,0,0,1\big)\, ,
\end{equation}
respectively are the SO(6)-invariant quadratic form corresponding to the electrically-gauged SO(6) factor of either gauge group (\ref{eq:GaugeGroups}), and the quadratic form of the magnetically gauged SO$(1,1)$ (if $x=-1$) or SO$(2)$ (if $x=1$) factors in (\ref{eq:GaugeGroups}). The gaugings $G_\pm$ (\ref{eq:GaugeGroups}) uplift consistently on $S^5 \times S^1$ to type IIB supergravity \cite{Inverso:2016eet} (see also \cite{Berman:2021ynm}). In particular, the vacua of these $D=4$ $\cN=8$ supergravities give rise to ten-dimensional solutions of the form (\ref{eq:AdS4ixS5xS1}). Upon uplift, the SO(6) factor of $G_\pm$ rotates $S^5$ while SO$(1,1)$ or SO$(2)$ act on the $S^1$. 

The vacua of $D=4$ $\cN=8$ supergravity with $G_\pm$ gaugings, all of which are known to be AdS, tend to come in families parameterised by supergravity moduli. Consequently, the resulting type IIB uplifts (\ref{eq:AdS4ixS5xS1}) also depend on these moduli. In this paper, we will be interested in three specific such families of AdS solutions. All of them preserve $\cN=1$ supersymmetry and have residual bosonic, continuous symmetry groups $G$. While $G$ for a given family possibly depends on the position in moduli space, it always turns out to be contained in the SU(3) subgroup of the $\textrm{SO}(6) \sim \textrm{SU}(4)$ factor of either gauge group (\ref{eq:GaugeGroups}). The full (super)symmetry group of these solutions, $\textrm{OSp}(4|1) \times G$, at a specific point in their moduli spaces is thus always contained in 
\begin{equation} \label{eq:Fullsusy}
\textrm{OSp}(4|1) \times \textrm{SU}(3) \; .
\end{equation}
The factor of $\textrm{OSp}(4|1)$ is common to all solutions and $G \subset \textrm{SU}(3) \subset \textrm{SU}(4) \sim \textrm{SO}(6) \subset G_\pm $. From the ten-dimensional perspective (\ref{eq:AdS4ixS5xS1}), $G$ acts on the $\mathbb{CP}^2$ base of the Hopf fibration of $S^5$.

The AdS vacua under consideration may have additional bosonic discrete symmetries. In fact, at least one representative in each family enjoys the same specific discrete symmetries that allow for a truncation of $G_\pm$-gauged $D=4$ $\cN=8$ supergravity to the $\cN=1$ seven-chiral model with scalar manifold 
\begin{equation} \label{eq:7Chirals}
	\Big[\frac{\rm SL(2,\mathbb{R})}{\rm SO(2)}\Big]^7 \; ,
\end{equation}
associated to the maximal torus of $\textrm{E}_{7(7)}/\textrm{SU}(8)$. In other words, at least one representative in each family of solutions can be recovered as a solution within the $\cN=1$ subtruncation of maximal supergravity specified by (\ref{eq:7Chirals}).
 In the next three subsections, we will denote the real scalar components of these multiplets by $(\varphi_i,\chi_i)$, $i=1,\dots,7$, with $\varphi_i$ proper scalars and $\chi_i$ pseudoscalars taking values on (\ref{eq:7Chirals}). We follow the parameterisation conventions of~\cite{Guarino:2020gfe}.

\newpage 
\subsection{Two-parameter $\textrm{U}(1)^2$-invariant solution with SU(3) enhancement}	\label{sec:U(1)2toSU3}

The first family of AdS solutions that we will consider occurs in the $G_-$ gauging and was first found in \cite{Guarino:2019oct}. The solution depends on two pseudoscalar parameters and preserves $\cN=1$ supersymmetry for all their values. At generic points in this two-dimensional moduli space, the bosonic symmetry of the solution is the $G \equiv \textrm{U}(1)^2 \equiv \textrm{U}(1)_1 \times \textrm{U}(1)_2$ Cartan subgroup of the SU(3) contained in the SO(6) factor of the gauge group $G_-$, namely,
\begin{equation} \label{eq:Sol1Branching}
\textrm{SO}(6) \sim \textrm{SU}(4) 
\supset 
\textrm{SU}(3) \times \textrm{U}(1)_\tau
\supset 
\textrm{SU}(2) \times \textrm{U}(1)_2 \times \textrm{U}(1)_\tau
\supset 
\textrm{U}(1)_1 \times \textrm{U}(1)_2 \times \textrm{U}(1)_\tau  \; .
\end{equation}
Here, $\textrm{U}(1)_1 \subset \textrm{SU}(2)$. The $\textrm{U}(1)_\tau$ that commutes with SU(3) inside $\textrm{SU}(4) \sim \textrm{SO}(6)$ is broken by the family of vacua, but is included in (\ref{eq:Sol1Branching}) for clarity.

The sector invariant under this U$(1)^2$ of $G_-$--gauged $D=4$ $\cN=8$ supergravity was studied in \cite{Guarino:2019oct}, and the two-parameter family of AdS vacuum solutions that we are interested in was already reported in that reference. In \cite{Guarino:2020gfe}, the same two-parameter family of vacua was found to be included in the seven-chiral sector (\ref{eq:7Chirals}) of the maximal supergravity. In our conventions and with $g=m=1$ in (\ref{eq: embtensorcomponents}) without loss of generality, the location of this family within the seven-chiral model occurs at
\begin{eqnarray}	\label{eq:vevsSU3family}
&	
	x=-1 \; , \qquad 
	e^{2\varphi_1}=e^{2\varphi_2}=e^{2\varphi_3}=\tfrac95 \; , 		\qquad
	e^{2\varphi_4}=e^{2\varphi_5}=e^{2\varphi_6}=e^{2\varphi_7}=\tfrac65\,,		\nonumber \\
&
	\chi_1+\chi_2+\chi_3=0 \; ,		\qquad
	 \chi_4 =  \chi_5 =  \chi_6 =  \chi_7 = - \tfrac{1}{\sqrt{6}} \,.
\end{eqnarray}
The pseudoscalars $\chi_1$, $\chi_2$ are free and parameterise the family of vacua, while $\chi_3$ is fixed in terms of them as indicated in (\ref{eq:vevsSU3family}). The squared radius of AdS on this family is independent of $\chi_1$, $\chi_2$ and reads, in our conventions,
\begin{equation}	\label{eq: SU3CC}
	L^2=\frac{-6}{V_0}=\frac{25\sqrt5}{108}\, ,
\end{equation}
with $V_0 <0$ the value of the cosmological constant at the critical locus (\ref{eq:vevsSU3family}).

While the family of vacua (\ref{eq:vevsSU3family}) preserves $\cN=1$ supersymmetry for all values of $\chi_1$, $\chi_2$, its bosonic symmetry depends on the position in moduli space. For generic values of the parameters, the solution is invariant under the $\textrm{U}(1)_1 \times \textrm{U}(1)_2$ specified in (\ref{eq:Sol1Branching}). On the one-dimensional locus $\chi_1 = \pm \chi_2$, there is a symmetry enhancement to either $G\equiv \textrm{SU}(2) \times \textrm{U}(1)_2$ or $G\equiv \textrm{U}(1)_1 \times \textrm{SU}(2)$ depending on whether $\chi_1 =  \chi_2$ or $\chi_1 =  - \chi_2$. Of course, both choices are physically equivalent. Finally, at the $\chi_1 = \pm \chi_2 = 0$ point, the symmetry enhances itself to $G \equiv \textrm{SU}(3)$. The full (super)symmetry group of the family is thus indeed contained in (\ref{eq:Fullsusy}), and becomes exactly the latter at the origin (in our conventions) of moduli space.  The $\chi_1 = \chi_2 = 0$, SU(3)-invariant solution was uplifted to type IIB in \cite{Guarino:2019oct}, and the two-parameter family was uplifted in \cite{Guarino:2021kyp}. See appendix \ref{sec:UpliftSU3Fam} below for a follow-up on the latter uplift.

As $D=4$ supergravity scalars, $\chi_1$, $\chi_2$ take values on two copies, contained in (\ref{eq:7Chirals}),  of the real line. However, on the associated family of type IIB solutions, a periodicity on these parameters must be enforced,
\begin{equation} \label{eq:periodicity}
\chi_1 \sim \chi_1 + \tfrac{2\pi}{T} \; , \qquad 
\chi_2 \sim \chi_2 + \tfrac{2\pi}{T} \; ,
\end{equation}
so the moduli space becomes a two-torus. The quantity $T$ in (\ref{eq:periodicity}) is the inverse radius of the S-folded $S^1$ of the corresponding type IIB solutions (\ref{eq:AdS4ixS5xS1}). This periodicity cannot be seen at the $D=4$ $\cN=8$ supergravity level and is an intrinsic feature of the uplifted type IIB solutions. This can be verified with the explicit uplift of appendix \ref{sec:UpliftSU3Fam}, and reflects itself in the KK spectrum discussed in section \ref{sec:U(1)2toSU3KK}. Similar observations about the compactness of the pseudoscalar moduli in related models have been previously made in \cite{Giambrone:2021zvp,Cesaro:2021tna,Giambrone:2021wsm}.

\subsection{One-parameter $\textrm{U}(1)$-invariant solution with SO(3) enhancement}	\label{sec:U(1)toSO3}

In \cite{Berman:2021ynm} (see also \cite{Bobev:2021rtg}), an AdS vacuum of the $G_+$ gauging (\ref{eq:GaugeGroups}) of $D=4$ $\cN=8$ supergravity was found within the seven chiral subsector (\ref{eq:7Chirals}). In our conventions, this vacuum is located at
\begin{eqnarray} \label{eq:sol2}
& x = 1 \; , \qquad e^{2\varphi_1}=e^{2\varphi_2}=e^{2\varphi_3}=\frac{27}{5}\; , \qquad e^{4\varphi_4}=e^{4\varphi_5}=e^{4\varphi_6}=e^{4\varphi_7}=\frac{108}{25} \; , \nonumber \\
& \chi_1=\chi_2=\chi_3=-\frac{2}{3\sqrt3} \;  , \qquad
  \chi_4=\chi_5=\chi_6=\frac{1}{108^{\nicefrac14}}\;  , \qquad
  \chi_7=-\frac{5}{108^{\nicefrac14}}\;  ,
\end{eqnarray}
and has AdS squared radius
\begin{equation} \label{eq:CCsol2}
L^2= \frac{-6}{V_0}= \frac{25 \sqrt{\frac{5}{3}}}{162} \; . 
\end{equation}
This solution preserves the SO(3) subgroup of the $G_+$ gauge group that is embedded in the SO(6) factor of the latter as
\begin{equation} \label{eq:Sol2Branching}
\textrm{SO}(6) \sim \textrm{SU}(4) 
\supset 
\textrm{SU}(3) \times \textrm{U}(1)_\tau
\supset 
\textrm{SO}(3) \; , 
\end{equation}
with $\textrm{SO}(3)$ the real subgroup of $\textrm{SU}(3)$ so that the fundamental is irreducible. 

The critical point (\ref{eq:sol2}) can be extended into a one-parameter family of AdS solutions of the same $D=4$ $\cN=8$ gauged supergravity \cite{Berman:2021ynm} (see also \cite{Bobev:2021rtg}). This can be done using the solution generating technique first introduced in \cite{Bobev:2021yya} and systematised in \cite{Guarino:2021hrc}. The modulus, $\chi$, that labels the family is a pseudoscalar of the $D=4$ $\cN=8$ supergravity that is not contained in (\ref{eq:7Chirals}). For all $\chi$ does the family have the same cosmological constant (\ref{eq:CCsol2}) and preserves $\cN=1$ supersymmetry. For generic $\chi$, the family only preserves the U(1) Cartan subgroup of the SO(3) in (\ref{eq:Sol2Branching}). At $\chi = 0$, the critical point (\ref{eq:sol2}) is attained and the symmetry is enhanced to SO(3). Only this $\chi = 0$ point can be recovered within the seven-chiral sector (\ref{eq:7Chirals}) of the $D=4$ supergravity. 

The $\chi=0$, SO(3)-invariant solution was uplifted in \cite{Berman:2021ynm} to a type IIB S-fold of the form (\ref{eq:AdS4ixS5xS1}). From the KK spectrum discussed in section \ref{sec:U(1)toSO3KK}, it follows that the type IIB uplift of this family of solutions must have the pseudoscalar parameter $\chi$ periodically identified as
\begin{equation} \label{eq:periodicitySO3Fam}
\chi \sim \chi + \tfrac{2\pi}{T} \; , 
\end{equation}
similarly to all other known cases, (\ref{eq:periodicity}) and \cite{Giambrone:2021zvp,Cesaro:2021tna}. In appendix \ref{sec:UpliftSU3Fam} we comment on the type IIB uplift of this family of AdS$_4$ solutions.

\newpage

\subsection{One-parameter $\textrm{U}(1)$-invariant solution with no enhancement}	\label{sec:U(1)}

A final one-parameter family of $\cN=1$ AdS vacua that we will consider was found in \cite{Bobev:2021rtg} within the $G_+$ gauging (\ref{eq:GaugeGroups}). The representative of this family contained in the seven chiral subsector (\ref{eq:7Chirals}) of the $D=4$ $\cN=8$ supergravity is located, in our conventions, at
\begin{equation}	\label{eq: u1solution}
	\begin{aligned}
		& x=1 \; , \quad \chi_1=-\frac{\kappa}{2\sqrt{6}}\,,
		\qquad
		\chi_2=\chi_3=-\frac{\sqrt{\kappa^2+3}}{4\sqrt2}\,,	
		\qquad
		\chi_4=-\frac{\kappa-2}{2\sqrt{3}(2\kappa^2-5)^{\nicefrac14}}\,,	
		\\[5pt]
		&\chi_5=\chi_6=\frac{\sqrt{\kappa}}{2\cdot 6^{\nicefrac14}}\,,	
		\qquad
		\chi_7=-\frac{\sqrt{\kappa^2+2}(2\kappa^2-5)^{\nicefrac14}}{\sqrt6(2-\kappa)}\,,
		\\[5pt]
		&e^{-\varphi_1}=\frac{\kappa}{2\sqrt2}\,,
		\qquad
		e^{-\varphi_2}=e^{-\varphi_3}=\frac{\sqrt{3\kappa^2-3}}{4\sqrt2}\,,	
		\qquad
		e^{-\varphi_4}=e^{-\varphi_7}=\frac{1}{\kappa}\Big(\frac{7\kappa^2}2-8\Big)^{\nicefrac14}\,,
		\\[5pt]
		&e^{-\varphi_5}=e^{-\varphi_6}=\Big(\frac3{32}\Big)^{\nicefrac14}\sqrt{\kappa}\,,
	\end{aligned}
\end{equation}
with $\kappa \equiv \sqrt{\sqrt{13}-1}$. Using (\ref{eq: u1solution}) as a seed for the solution generating technique of \cite{Bobev:2021yya,Guarino:2021hrc}, a one-parameter family of AdS solutions parameterised by a pseudoscalar $\chi$ can be obtained \cite{Bobev:2021rtg}. All members of the family preserve $\cN=1$ supersymmetry and the same $G \equiv \textrm{U}(1)$ bosonic symmetry as the family discussed in section \ref{sec:U(1)toSO3}, namely the Cartan subgroup of the SO(3) in (\ref{eq:Sol2Branching}). Also, all members of the family share the following $\chi$-independent AdS squared radius:
\begin{equation}
		L^2= \frac{-6}{V_0}= \frac{81}{32\sqrt{70+26\sqrt{13}}}\, .
\end{equation}
The modulus $\chi$ is a $D=4$ $\cN=8$ supergravity scalar that lies outside the seven chiral subsector (\ref{eq:7Chirals}). For this reason, only the $\chi = 0$ representative (\ref{eq: u1solution}) of this family of vacua lies within that sector. Unlike in the previous cases, there is no symmetry enhancement anywhere in the family, including $\chi=0$. Like in the previous solutions, the parameter $\chi$ is restricted to take values on a circle in the full type IIB solution, as in (\ref{eq:periodicitySO3Fam}). This observation follows from the KK spectrum for this solution covered in section \ref{sec:U(1)KK}.


\section{Spectra} \label{sec:spectra}


Let us now move on to present the KK spectrum for each of the families of AdS$_4$ solutions reviewed in section \ref{sec:AdSSols}, starting with aspects common to all of them.

\subsection{Common aspects} \label{sec:CommonAspects}

The characterisation of the KK spectrum on the AdS$_4$ solutions at hand can be achieved by, firstly, obtaining the structure of $\textrm{OSp}(4|1) \times G$ representations and, secondly, computing the $\textrm{OSp}(4|1)$ supermultiplet dimensions.

Finding the complete algebraic structure of OSp$(4|1)$ multiplets and their charges under the (parameter-dependent) bosonic symmetry group $G$ is a purely group-theoretical exercise. This structure is dictated by the consitent uplift and follows from an adaptation to the present setting of the technique employed in \cite{Englert:1983rn} in a related context. The structure of individual KK states follows from tensoring the $D=4$ $\cN=8$ supergravity multiplet with the $[0,\ell,0]_{2n}$ representation of $\textrm{SO}(6) \times \textrm{SO}(2)$ for all $\ell$ and $n$ ranging as
\begin{equation} \label{eq:KKlevels}
\ell = 0, 1 , 2 , \ldots \qquad n = 0 , \, \pm 1 ,  \, \pm 2 , \, \ldots 
\end{equation}
These so-called KK levels are associated with the internal $S^5$ and $S^1$ of the IIB S-folds (\ref{eq:AdS4ixS5xS1}). The resulting $\textrm{SO}(6) \times \textrm{SO}(2)$ representations can be found in appendix B of \cite{Cesaro:2021tna}. These then need to be branched under $G$ and regrouped into OSp$(4|1)$ multiplets with the same $G$ charges. 

As remarked in section~\ref{sec:AdSSols} above, the (super)symmetry group $\textrm{OSp}(4|1) \times G$ for all three families of solutions under consideration in this paper is contained in (\ref{eq:Fullsusy}). Consequently, the supermultiplets present in all three spectra must branch from putative representations of the latter. For this reason, it is helpful to present these putative representations in detail. These are obtained by tensoring as outlined above, then branching under the first inclusion in (\ref{eq:Sol1Branching}), and finally recombining into OSp$(4|1)$ supermultiplets with the same SU(3) charges. Denoting the representations of (\ref{eq:Fullsusy}) as\footnote{Here, the acronym MULT refers to a supermultiplet of OSp$(4|1)$ with dimension $E_0$ and any possible superconformal primary spin $s_0<2$: $s_0 = \frac32$ ((M)GRAV), $s_0 = 1$ (GINO), $s_0 = \frac12$ ((M)VEC) or $s_0=0$ (CHIRAL), where M denotes a massless multiplet. The Dynkin labels $[p,q]$ characterise the SU(3) representation. All these multiplets are long, except the massless multiplets MGRAV and MVEC, for which $E_0 = \frac52$ and $E_0 = \frac32$, respectively. See {\it e.g.}~table 1 of \cite{Cesaro:2020soq} for their field content. No massless GINO  or singleton multiplets appear in the spectra.} 
\begin{equation}	\label{eq: notationSU3point}
	{\rm MULT}[E_0;\, [p,q]] \; , 
\end{equation}
we find the following content at  KK level $\ell = 0$ and all $n$:
\begin{align} \label{eq:AllSU3Mults_ell0}
	&{\rm \underline{GRAV} }\big[E_0;\, [0,0]\big]	
	\oplus{\rm GINO}\big[E_0;\, [1,0]\big]\oplus{\rm GINO}\big[E_0;\, [0,1]\big]		\nonumber\\[5pt]
	&\quad\oplus{\rm \underline{VEC}}\big[E_0;\, [1,1]\big]\oplus{\rm VEC}\big[E_0;\, [1,0]\big]\oplus{\rm VEC}\big[E_0;\, [0,1]\big]	\nonumber\\[5pt]
	&\qquad\oplus{\rm CHIRAL}\big[E_0;\, [2,0]\big]\oplus{\rm CHIRAL}\big[E_0;\, [0,2]\big]\oplus2\times{\rm CHIRAL}\big[E_0;\, [0,0]\big] \; .
\end{align}
At fixed $\ell\geq1$ and for all $n$, the (\ref{eq:Fullsusy}) representation content is:
\begin{align} \label{eq:AllSU3Mults}
	&\bigoplus_{p=0}^{\ell}{\rm GRAV}\big[E_0;\, [p,\ell-p]\big]	\nonumber\\[5pt]
	&\oplus\bigoplus_{p=0}^{\ell}\bigoplus_{a=0}^{1}{\rm GINO}\big[E_0;\, [p+1-a,\ell-p+a]\big]		\nonumber\\[5pt]
	&\qquad\oplus\bigoplus_{p=0}^{\ell-1}\bigoplus_{a=0}^{1}\Big({\rm GINO}\big[E_0;\, [p+a,\ell-p-1]\big]
	\oplus{\rm GINO}\big[E_0;\, [p,\ell-p+a-1]\big]\Big)	\nonumber\\[5pt]
	&\oplus\bigoplus_{p=0}^{\ell}\bigoplus_{a=0}^{1}\bigoplus_{b=0}^{1}{\rm VEC}\big[E_0;\, [p+a,\ell-p+b]\big]		\nonumber\\[5pt]
	&\qquad\oplus\bigoplus_{p=0}^{\ell-1}\bigoplus_{a=0}^{2}\Big({\rm VEC}\big[E_0;\, [p+a,\ell-p-1]\big]
	\oplus{\rm VEC}\big[E_0;\, [p,\ell-p+a-1]\big]\Big)	\nonumber\\[5pt]
	&\qquad\qquad\oplus\bigoplus_{p=0}^{\ell-2}\bigoplus_{a=0}^{1}\bigoplus_{b=0}^{1}{\rm VEC}\big[E_0;\, [p+a,\ell-p+b-2]\big]		\nonumber\\[5pt]
	&\oplus\bigoplus_{p=0}^{\ell}\bigoplus_{a=0}^{1}\Big({\rm CHIRAL}\big[E_0;\, [p+2a,\ell-p]\big]
	\oplus{\rm CHIRAL}\big[E_0;\, [p,\ell-p+2a]\big]\Big)		\nonumber\\[5pt]
	&\qquad\oplus\bigoplus_{p=0}^{\ell-1}\bigoplus_{a=0}^{1}\Big({\rm CHIRAL}\big[E_0;\, [p+1-a,\ell-p]\big]
	\oplus{\rm CHIRAL}\big[E_0;\, [p+1,\ell-p-a]\big]\Big)	\nonumber\\[5pt]
	&\qquad\qquad\oplus\bigoplus_{p=0}^{\ell-2}\bigoplus_{a=0}^{2}\Big({\rm CHIRAL}\big[E_0;\, [p+a,\ell-p-2]\big]
	\oplus{\rm CHIRAL}\big[E_0;\, [p,\ell-p+a-2]\big]\Big)	\, .
\end{align}
For each solution with symmetry $G$ at a certain location in moduli space, the complete $\textrm{OSp}(4|1) \times G$ algebraic structure of the spectrum at all KK levels $\ell$ and $n$ (\ref{eq:KKlevels}) follows from branching the SU(3) representations in (\ref{eq:AllSU3Mults_ell0}), (\ref{eq:AllSU3Mults}) under $G\subset \textrm{SU}(3)$. The $\textrm{OSp}(4|1)$ multiplets are the same as in these equations, but their dimensions $E_0$ will typically also split under the branching $G\subset \textrm{SU}(3)$. The role of the underlined $\ell=0$ multiplets in (\ref{eq:AllSU3Mults_ell0}) will be discussed around (\ref{eq: shortening_generalGRAV}) and (\ref{eq: shortening_generalVEC}) below. 

The supermultiplet structure (\ref{eq:AllSU3Mults_ell0}), (\ref{eq:AllSU3Mults}) of the KK spectra of all three families is independent of $n$, and only depends on $\ell$ as indicated. The only dependence of the spectrum on $n$ is through the dimensions $E_0$, see below. The total multiplicities of each type of OSp$(4|1)$ multiplet at fixed KK levels is thus also independent of $n$. Naively adding up the dimensions of the different SU(3) representations in (\ref{eq:AllSU3Mults_ell0}), (\ref{eq:AllSU3Mults}) for each type of OSp$(4|1)$ multiplet at fixed KK level $\ell$, these multiplicities are
\begin{equation} \label{eq:GenericCounting}
\begin{tabular}{cclcccl}
$\textrm{GRAV}$  &:&  $D_{\ell,6}$ \; , &\qquad \qquad &
$\textrm{GINO}$  &:&  $6 D_{\ell,6}$ \; , 		\\
$\textrm{VEC}$  &:&  $14 D_{\ell,6}$ 	\; , &\qquad \qquad &
$\textrm{CHIRAL}$  &:&  $14 D_{\ell,6}$ \; , 
\end{tabular}
\end{equation}
for all $n$. Here,
\begin{equation}
	D_{\ell,6} \equiv {{\ell+5}\choose{\ell}}-{{\ell+3}\choose{\ell-2}}=\tfrac1{12}(\ell+3)(\ell+2)^2(\ell+1)
\end{equation}
is the dimension of the $[0,\ell,0]$ representation of $\textrm{SU}(4) \sim \textrm{SO}(6)$. This is a consequence of the fact that all multiplets can be taken to be long, and of the generic algebraic structure of the KK spectrum on the class of AdS$_4$ solutions (\ref{eq:AdS4ixS5xS1}) of type IIB that we are considering. 

Having obtained the algebraic structure of the spectra, it remains to determine the supermultiplet dimensions. Except for the handful of massless cases mentioned below, the OSp$(4|1)$ multiplets present in (\ref{eq:AllSU3Mults_ell0}), (\ref{eq:AllSU3Mults}) are all typically long. For this reason, the dimensions do not follow from group theory and must be computed independently. In order to do this, we have used the mass matrices given in \cite{Malek:2019eaz,Malek:2020yue,Cesaro:2020soq} (see also \cite{Varela:2020wty}) for the KK perturbations above AdS backgrounds that uplift from $D=4$ $\cN=8$ gauged supergravities. As remarked in those references, only gauged supergravity data together with minimal information on the uplift, including the generators of the maximal isometry $\textrm{SO}(6) \times \textrm{SO}(2)$ of $S^5 \times S^1$, enter those mass matrices. Using the details reviewed in section \ref{sec:AdSSols} and the generators of  $\textrm{SO}(6) \times \textrm{SO}(2)$ (see appendix B of \cite{Cesaro:2021tna} for our conventions), we have evaluated these mass matrices for each of these families of solutions. Finally, we have diagonalised them at some fixed KK levels to obtain the mass spectrum of the individual KK perturbations at those levels. Then, we have translated these masses into dimensions, and have grouped them  into OSp$(4|1)$ supermultiplets. It is reassuring that the latter step agrees, as it must, with the independent algebraic supermultiplet structure (\ref{eq:AllSU3Mults_ell0}), (\ref{eq:AllSU3Mults}).

Our results for the multiplet dimensions are mostly numerical, and allow us to infer important patterns. We have also obtained some analytic results, which confirm those patterns. Specifically, the supermultiplet dimensions $E_0$ typically exhibit a complicated dependence on the $S^5$ KK level $\ell$ and possibly on the quantum numbers of intermediate, broken symmetry groups. However, their dependence on the $S^1$ KK level $n$, on the moduli $(\chi_1 , \chi_2)$ or $\chi$, and on the $G= \textrm{U}(1)_1 \times \textrm{U}(1)_2$ or $G = \textrm{U}(1)$ charges  $(m_1 , m_2)$ or $m$ is always locked into the fixed combinations
\begin{equation} \label{eq:chiComb}
f_{nm_1m_2} (\chi_1 , \chi_2) \equiv \big[ \tfrac{2\pi n}{T} + \tfrac12 m_1 (\chi_1 - \chi_2) + \tfrac12 m_2 (\chi_1 + \chi_2) \big]^2  \qquad
\textrm{or}  \qquad
\big( \tfrac{2\pi n}{T}+ m \chi \big)^2 \; .
\end{equation}
The expression on the left corresponds to the two-parameter family of \cite{Guarino:2019oct,Guarino:2020gfe} that was reviewed in section \ref{sec:U(1)2toSU3}, and the one on the right to the one-parameter families of  \cite{Berman:2021ynm,Bobev:2021rtg} reviewed in sections \ref{sec:U(1)toSO3} and \ref{sec:U(1)}. In (\ref{eq:chiComb}), $T$ is the inverse radius of the S-folded $S^1$ of the corresponding type IIB solutions. A similar behaviour has been observed for the spectra discussed in \cite{Giambrone:2021zvp,Cesaro:2021tna}.

The behaviour (\ref{eq:chiComb}) has various consequences. Firstly, it follows that an OSp$(4|1)$ multiplet in each spectrum is neutral under $\textrm{U}(1)_1 \times \textrm{U}(1)_2$ or $\textrm{U}(1)$ if and only if its dimension is independent of the moduli that parameterise the family. Secondly, (\ref{eq:chiComb}) also leads to degeneracy in $E_0$ of the multiplets at $S^1$ KK levels $|n|$ and $-|n|$, at fixed moduli and $S^5$ KK level $\ell$, and opposite  $\textrm{U}(1)_1 \times \textrm{U}(1)_2$ or $\textrm{U}(1)$ charges. Finally, equation (\ref{eq:chiComb}) also establishes the periodic behaviour of the multiplet dimensions in the moduli. Indeed, for all fixed $S^5$ KK level $\ell$, the dimension of any given multiplet with charges $(m_1 , m_2)$, evaluated at $(\chi_1 = \chi_{1\0} , \chi_2 = \chi_{2\0})$ or $\chi = \chi_{\0}$ and $S^1$ KK level $n$, coincides with the dimension of the same multiplet evaluated at $( \chi_1 = \chi_{1\0} + \frac{2\pi h_1}{T} , \chi_2 = \chi_{2\0} + \frac{2\pi h_2}{T} )$ or $\chi = \chi_\0 + \frac{2\pi}{T}$ and $S^1$ level $n^\prime$, with 
\begin{equation} \label{eq:KKreshuffle}
n^\prime = n - \tfrac12 (m_1 + m_2 ) \, h_1 +  \tfrac12 (m_1 -  m_2 ) \, h_2  \qquad
\textrm{or}  \qquad
n^\prime = n - m \;.
\end{equation}
Independent integer winding numbers $(h_1 , h_2 )$ need to be introduced in the expressions corresponding to the family of solutions of section \ref{sec:U(1)2toSU3}, but are unnecessary for the solutions of sections \ref{sec:U(1)toSO3} and \ref{sec:U(1)}. Either way, the KK level $n^\prime$ in (\ref{eq:KKreshuffle}) is well defined, as all charges $m_1$, $m_2$ and $m$ are integer in our conventions, and both $(m_1 + m_2 )$ and $(m_1 - m_2 )$ are even numbers: see (\ref{eq:SU(3)toCartans}). The periodicity of the multiplet dimensions, together with the $n$-independence of the algebraic structure (\ref{eq:AllSU3Mults_ell0}), (\ref{eq:AllSU3Mults}), leads to the periodicity in moduli space of the full spectra.

All the multiplets (\ref{eq:AllSU3Mults}) present in the spectra of all three families at KK levels $\ell \geq 1$ and any $n$ are always long. At KK level $\ell =0$, most of the multiplets (those not underlined) in (\ref{eq:AllSU3Mults_ell0}) are also long, but the underlined singlet GRAV and $\bf{8}$ VECs therein behave differently. For $n \neq 0$, this GRAV multiplet is long as indicated, but at $n=0$ it becomes massless for all values of the parameters $(\chi_1 , \chi_2)$ or $\chi$ via the splitting:
\begin{equation}	\label{eq: shortening_generalGRAV}
		{\rm \underline{GRAV}}[E_0=\tfrac52+\epsilon] \xrightarrow[\epsilon\to0]{}{\rm MGRAV}[E_0=\tfrac52]\oplus{\rm GINO}[E_0=3] \; .
\end{equation}
Only the dimensions, and not any other (vanishing) charges, are shown here. The MGRAV multiplet on the r.h.s.~of (\ref{eq: shortening_generalGRAV}) contains the $\ell = n =0$ massless graviton of the $D=4$ $\cN=8$ gauged supergravity, and is accompanied by an additional massive GINO multiplet of fixed dimension, as indicated in (\ref{eq: shortening_generalGRAV}). 

As for the underlined $\ell =0$ VEC multiplets in (\ref{eq:AllSU3Mults_ell0}), a number $\textrm{dim} \, G$ of these turn out to lie in the adjoint of $G$ under the branching of the $\bm{8}$ of SU(3) under $G$, for all possible symmetry groups $G$. These $\textrm{dim} \, G$ VEC multiplets must be replaced with
\begin{equation}	\label{eq: shortening_generalVEC}
		{\rm \underline{VEC}}[E_0=\tfrac32+\epsilon] \xrightarrow[\epsilon\to0]{}{\rm MVEC}[E_0=\tfrac32]\oplus{\rm CHIRAL}[E_0=2] \; .
\end{equation}
Unlike the straightforward replacement (\ref{eq: shortening_generalGRAV}) at $n =0$ across the entire moduli spaces, the splitting (\ref{eq: shortening_generalVEC}) is slightly more subtle. At generic loci with minimal symmetry, $G = \textrm{U}(1)^2$ for the family of \cite{Guarino:2019oct,Guarino:2020gfe} and  $G = \textrm{U}(1)$ for the families of \cite{Berman:2021ynm,Bobev:2021rtg}, (\ref{eq: shortening_generalVEC}) does work out like (\ref{eq: shortening_generalGRAV}): the replacement takes place only at $n=0$ and the VEC multiplets stay massive for $n \neq 0$. In these cases, the multiplets on the r.h.s.~of (\ref{eq: shortening_generalVEC}) respectively contain the U$(1)^2$ or U(1) gauge fields and the massless moduli $(\chi_1 , \chi_2)$ or $\chi$, all of which are $D=4$ $\cN=8$ supergravity fields. At loci or points with enhanced symmetry $G = \textrm{SU}(2) \times \textrm{U}(1)$, $G = \textrm{SU}(3)$ or $G = \textrm{SO}(3)$, the replacement (\ref{eq: shortening_generalVEC}) must be effected for all the VEC multiplets in the adjoint of $G$. In these cases, (\ref{eq: shortening_generalVEC}) occurs at either $n=0$ or at an $n^\prime \neq 0$ given by (\ref{eq:KKreshuffle}), depending on whether the symmetry enhancements occur at the locations specified in $D=4$ $\cN=8$ supergravity ($\chi_1 = \pm \chi_2$, $\chi_1 = \pm \chi_2 = 0 $ or $\chi=0$), or at locations periodically identified with the former via (\ref{eq:periodicity}) or (\ref{eq:periodicitySO3Fam}).  As noted in \cite{Giambrone:2021zvp,Cesaro:2021tna} in a similar context, the $n^\prime \neq 0$ situation is reminiscent of the `space invaders scenario' described in \cite{Duff:1986hr} (see also \cite{Cesaro:2020piw}). 

Whenever the replacements (\ref{eq: shortening_generalGRAV}), (\ref{eq: shortening_generalVEC}) occur, the generic supermultiplet degeneracies  (\ref{eq:GenericCounting}) need to be adapted to account for the extra GINOs and CHIRALs. At symmetry-enhanced points, these CHIRALs contain additional massless scalars. These are not moduli, however, since they become massive as $(\chi_1 , \chi_2)$ or $\chi$ move away from the symmetry-enhanced locations into generic loci with U$(1)^2$ or U(1) symmetry. Incidentally, our analysis seems to suggest that $(\chi_1 , \chi_2)$ and $\chi$ are the only moduli of the families of solutions under consideration, but this is not conclusive\footnote{Our numerics reveal massless scalars at higher KK levels, but these are likely to be an artifact of our choice $T=2\pi$ for the inverse radius of the S-folded $S^1$. See \cite{Assel:2018vtq,Berman:2021ynm} for more realistic, physically motivated choices for $T$.}. Finally, no GINO multiplet ever becomes massless. Thus, the supersymmetry stays strictly $\cN=1$ across the moduli spaces of all three families.

\newpage


\subsection{Two-parameter $\textrm{U}(1)^2$-invariant solution with SU(3) enhancement}	\label{sec:U(1)2toSU3KK}

For the family of solutions of \cite{Guarino:2019oct,Guarino:2020gfe} reviewed in section \ref{sec:U(1)2toSU3}, we have been able to determine the multiplet dimensions in closed form at all KK levels and for all values of the parameters $\chi_1$, $\chi_2$. Together with the full supermultiplet content that we give in all cases, this determines the complete spectrum for this family in full detail. Our results contain and extend previous partial results, including the graviton spectrum at the SU(3) point \cite{Dimmitt:2019qla} and the $\ell = n = 0$ spectrum on the two-parameter family \cite{Guarino:2020gfe}.

Let us first discuss the complete spectrum for generic values of the parameters $\chi_1$, $\chi_2$ away from symmetry-enhanced points. At fixed levels $\ell$ and $n$ ranging as in (\ref{eq:KKlevels}), the KK spectrum is arranged into the following representations of $\textrm{OSp}(4|1) \times \textrm{U}(1)_1 \times \textrm{U}(1)_2$. The OSp$(4|1)$ multiplets are those that appear in (\ref{eq:AllSU3Mults_ell0}), (\ref{eq:AllSU3Mults}), and their $\textrm{U}(1)_1 \times \textrm{U}(1)_2$ charges $(m_1 , m_2)$ branch from the SU(3) representations $[p,q]$ therein under $\textrm{U}(1)_1 \times \textrm{U}(1)_2 \subset \textrm{SU}(3)$, namely,
\begin{equation} \label{eq:SU(3)toCartans}
	[p,q]\to\bigoplus_{a=0}^{p}\bigoplus_{b=0}^{q}\bigoplus_{m=0}^{a+b}\big(a+b-2m,\ 2(q-p)+3(a-b)\big) \, .
\end{equation}
The supermultiplet dimensions depend on the KK levels, on these U$(1)^2$ charges and on the putative SU(3) Dynkin labels. More concretely, the dimension $E_0$ of each  OSp$(4|1)$ multiplet with superconformal primary spin $s_0$, present in the spectrum at KK levels $\ell$, $n$, with U$(1)^2$ charges $(m_1 , m_2)$ that derive through (\ref{eq:SU(3)toCartans}) from the SU(3) representations $[p,q]$ indicated in (\ref{eq:AllSU3Mults_ell0}), (\ref{eq:AllSU3Mults}), is
\begin{equation} \label{eq:GenericE0}
E_0=1+\sqrt{6 -s_0(s_0 +1) +\tfrac{5}{4} \ell(\ell+4) -\tfrac{5}{3}\mathcal{C}_2(p,q) + \tfrac54 f_{nm_1m_2} (\chi_1 , \chi_2) } \; .
\end{equation}
Here, $\mathcal{C}_2(p,q)  \equiv \frac13[p(p+3)+q(q+3)+pq]$ is the eigenvalue of the quadratic Casimir operator of SU(3) on the $[p,q]$ representation, and $f_{nm_1m_2} (\chi_1 , \chi_2)$ has been defined in (\ref{eq:chiComb}). Specifically, the dimension of each possible type of OSp$(4|1)$ multiplet present in the spectrum is 
\begin{equation}	\label{eq: conformaldimensions}
	\begin{tabular}{rcl}
		(M)GRAV	&:&	$E_0=1+\sqrt{\tfrac94+\tfrac{5}{4} \ell(\ell+4) -\tfrac{5}{3}\mathcal{C}_2(p,q) + \tfrac54 f_{nm_1m_2} (\chi_1 , \chi_2) }$\,,		\\[7pt]
		GINO	&:&	$E_0=1+\sqrt{4+\tfrac{5}{4} \ell(\ell+4) -\tfrac{5}{3}\mathcal{C}_2(p,q) + \tfrac54 f_{nm_1m_2} (\chi_1 , \chi_2) }$\,,			\\[7pt]
		(M)VEC	&:&	$E_0=1+\sqrt{\tfrac{21}4+\tfrac{5}{4} \ell(\ell+4)-\tfrac{5}{3}\mathcal{C}_2(p,q) + \tfrac54 f_{nm_1m_2} (\chi_1 , \chi_2) }$\,,		\\[7pt]
		CHIRAL	&:&	$E_0=1+\sqrt{6+\tfrac{5}{4} \ell(\ell+4)-\tfrac{5}{3}\mathcal{C}_2(p,q) + \tfrac54 f_{nm_1m_2} (\chi_1 , \chi_2) }$\, . 			
	\end{tabular}
\end{equation}

As (\ref{eq:GenericE0}) or (\ref{eq: conformaldimensions}) show, only those multiplets charged under $\textrm{U}(1)_1 \times \textrm{U}(1)_2$ have their dimensions depend on the parameters $\chi_1$ and $\chi_2$, in agreement with the general discussion of section \ref{sec:CommonAspects}. The underlined $\ell=0$ GRAV multiplet in (\ref{eq:AllSU3Mults_ell0}) is massive for $n \neq 0$, but becomes massless through (\ref{eq: shortening_generalGRAV}) at $n=0$, for all values of $\chi_1$ and $\chi_2$. Similarly, the two U$(1)^2$ singlets that branch from the $\bm{8}$ underlined VEC multiplets at $\ell=0$ in (\ref{eq:AllSU3Mults_ell0}) are massive (at $n \neq 0$) or become massless (at $n=0)$ via (\ref{eq: shortening_generalVEC}) across the two-dimensional moduli space. These $\ell = n =0$ MVEC multiplets contain the vectors that gauge the residual U$(1)^2$ in the $D=4$ $\cN=8$ supergravity. The two accompanying CHIRAL multiplets resulting from (\ref{eq: shortening_generalVEC}) contain the massless moduli $\chi_1$ and $\chi_2$. For this family of solutions and for generic values of the parameters, there are no additional moduli in the spectrum.

In the locus $\chi_1 = \pm \chi_2$ or identifications (\ref{eq:periodicity}) thereof, the symmetry is enhanced to $\textrm{SU}(2) \times \textrm{U}(1)$. The complete spectrum on this locus is thus given by (\ref{eq:AllSU3Mults_ell0}), (\ref{eq:AllSU3Mults}), with the SU(3) representations $[p,q]$ therein branched out under the $\textrm{SU}(2) \times \textrm{U}(1)_2$ in (\ref{eq:Sol1Branching}) as
\begin{equation} \label{eq:SU3toSU2xU1}
	[p,q]\to\bigoplus_{a=0}^{p}\bigoplus_{b=0}^{q}\bm{\big(a+b+1\bm)}_{2(q-p)+3(a-b)} \; .
\end{equation}
Here, $\bm{\big(a+b+1\bm)}$ gives the dimension of the SU(2) representation, and the subindex, the U$(1)_2$ charge. The multiplet dimensions are still given by (\ref{eq:GenericE0}), now evaluated at $\chi_1 = \pm \chi_2$ modulo (\ref{eq:periodicity}). Finally, at the SU(3) invariant locations, the spectrum is given exactly by (\ref{eq:AllSU3Mults_ell0}), (\ref{eq:AllSU3Mults}), where the supermultiplet dimensions are again given by (\ref{eq:GenericE0}), now evaluated at $\chi_1 = \pm \chi_2 = 0 $ modulo (\ref{eq:periodicity}). In these symmetry enhanced cases, extra massless multiplets occur in the spectrum following the pattern discussed around equation (\ref{eq: shortening_generalVEC}). For the reader's convenience, we have tabulated the KK spectrum at the $\chi_1 = \chi_2 = 0$ SU(3)-invariant point in tables \ref{table: (l=0,n)multipletsSU(3)} (at $\ell =0$ and all $n$) and \ref{table: (l=1,n)multipletsSU(3)} (at $\ell =1$ and all $n$). The tables provide the ${\rm OSp(4\vert1)}\times$SU(3) representation content at these levels following the notation (\ref{eq: notationSU3point}), only with the SU(3) representations $[p,q]$ factored out in common cells. See appendix \ref{sec: lowKKSU3} for the explicit spectrum at KK levels $\ell =0$ and all $n$ across the two-parameter family of solutions.

\begin{table}[]
\begin{center}
{\footnotesize
\begin{tabular}{|p{50mm}|p{50mm}|p{45mm}|} \hline
$[0,0]$ 							& 	$[0,1]$ 			& 	$[0,2]$ 		 		\\
$\text{\underline{GRAV}}\Big[1+\sqrt{\tfrac{9}{4}+\tfrac{5}{4}(\tfrac{2\pi n}{T})^2}\Big]$ 						& 	 same as $[1,0]$	& 		same as $[2,0]$			 \\[6pt]
$2\times\text{CHIRAL}\Big[1+\sqrt{6+\tfrac{5}{4}\left(\tfrac{2\pi n}{T}\right)^2}\Big]$	                			& 	 	&  \\
						& 				 		& 						 \\ \hline
$[1,0]$ 											& 	$[1,1]$ 								 	\\ 
	$\text{GINO}\Big[1+\sqrt{\tfrac{16}{9}+\tfrac{5}{4}\left(\tfrac{2\pi n}{T}\right)^2}\Big]$			& 	$\text{\underline{VEC}}\Big[1+\sqrt{\tfrac{1}{4}+\tfrac{5}{4}\left(\tfrac{2\pi n}{T}\right)^2}\Big]$								\\[6pt]
		$\text{VEC}\Big[1+\sqrt{\tfrac{109}{36}+\tfrac{5}{4}\left(\tfrac{2\pi n}{T}\right)^2}\Big]$			& 										 \\[6pt]\cline{1-2}
$[2,0]$ 																		\\
$\text{CHIRAL}\Big[1+\sqrt{\tfrac{4}{9}+\tfrac{5}{4}\left(\tfrac{2\pi n}{T}\right)^2}\Big]$						 					\\
 \\ \cline{1-1}
\end{tabular}
}
\caption{ ${\rm OSp(4\vert1)}\times$SU(3) representations present in the KK spectrum at levels $\ell=0$ and all $n$ of the $\chi_1 = \chi_2 = 0$ SU(3)-invariant solution of \cite{Guarino:2019oct,Guarino:2020gfe} reviewed in section \ref{sec:U(1)2toSU3}. At $n=0$, the underlined multiplets split as in (\ref{eq: shortening_generalGRAV}), (\ref{eq: shortening_generalVEC}).}
\label{table: (l=0,n)multipletsSU(3)}
\end{center}
\end{table}

\begin{table}[H]
\begin{center}
{\footnotesize
\begin{tabular}{|p{50mm}|p{47mm}|p{20mm}|p{20mm}|} \hline
$[0,0]$ 							& 	$[0,1]$ 			& 	$[0,2]$ 	& $[0,3]$	 		\\
$2\times\text{GINO}\Big[1+\sqrt{\tfrac{41}{4}+\tfrac{5}{4}(\tfrac{2\pi n}{T})^2}\Big]$ 			& same as $[1,0]$	 	&	same as $[2,0]$	& same as $[3,0]$	 \\
$2\times\text{VEC}\Big[1+\sqrt{\frac{23}{2}+\tfrac{5}{4}\left(\tfrac{2\pi n}{T}\right)^2}\Big]$	            	& 	 	&   & \\
						& 				 		& 					&	 \\ \hline
$[1,0]$ 											& 	$[1,1]$ 			& $[1,2]$					 	\\ 
$\text{GRAV}\Big[1+\sqrt{\tfrac{113}{18}+\tfrac{5}{4}(\tfrac{2\pi n}{T})^2}\Big]$& 	$2\times\text{GINO}\Big[1+\sqrt{\tfrac{21}{4}+\tfrac{5}{4}\left(\tfrac{2\pi n}{T}\right)^2}\Big]$		& same as $[2,1]$						\\[6pt]
$\text{GINO}\Big[1+\sqrt{\tfrac{289}{36}+\tfrac{5}{4}(\tfrac{2\pi n}{T})^2}\Big]$		& 		$2\times\text{VEC}\Big[1+\sqrt{\tfrac{13}{2}+\tfrac{5}{4}\left(\tfrac{2\pi n}{T}\right)^2}\Big]$			&										 \\[6pt]
$2\times\text{VEC}\Big[1+\sqrt{\tfrac{167}{18}+\tfrac{5}{4}(\tfrac{2\pi n}{T})^2}\Big]$		& $	2\times\text{CHIRAL}\Big[1+\sqrt{\tfrac{29}{4}+\tfrac{5}{4}(\tfrac{2\pi n}{T})^2}\Big]$							&		 \\[6pt]
$3\times\text{CHIRAL}\Big[1+\sqrt{\tfrac{361}{36}+\tfrac{5}{4}(\tfrac{2\pi n}{T})^2}\Big]$		& 					&					 \\[12pt]\cline{1-3}
$[2,0]$ 			& $[2,1]$															\\
$\text{GINO}\Big[1+\sqrt{\tfrac{169}{36}+\tfrac{5}{4}(\tfrac{2\pi n}{T})^2}\Big]$		& 	$\text{VEC}\Big[1+\sqrt{\tfrac{47}{18}+\tfrac{5}{4}(\tfrac{2\pi n}{T})^2}\Big]$									 \\	[6pt]	
$2\times\text{VEC}\Big[1+\sqrt{\tfrac{107}{18}+\tfrac{5}{4}(\tfrac{2\pi n}{T})^2}\Big]$		& 	$\text{CHIRAL}\Big[1+\sqrt{\tfrac{121}{36}+\tfrac{5}{4}(\tfrac{2\pi n}{T})^2}\Big]$									 \\			& 					
 \\ \cline{1-2}
$[3,0]$ 																		\\
$\text{CHIRAL}\Big[1+\sqrt{\tfrac{9}{4}+\tfrac{5}{4}(\tfrac{2\pi n}{T})^2}\Big]$											 \\	[6pt]		 					
 \\ \cline{1-1}
\end{tabular}
}
\caption{ ${\rm OSp(4\vert1)}\times$SU(3) representations present in the KK spectrum at levels $\ell=1$ and all $n$ of the $\chi_1 = \chi_2 = 0$ SU(3)-invariant solution of \cite{Guarino:2019oct,Guarino:2020gfe} reviewed in section \ref{sec:U(1)2toSU3}.}
\label{table: (l=1,n)multipletsSU(3)}
\end{center}
\end{table}

%

\subsection{One-parameter $\textrm{U}(1)$-invariant solution with SO(3) enhancement}	\label{sec:U(1)toSO3KK}

We now move on to discuss the KK spectrum on the one-parameter family of AdS$_4$ solutions of \cite{Berman:2021ynm} (see also \cite{Bobev:2021rtg}), reviewed in section \ref{sec:U(1)toSO3}. At generic values of the parameter $\chi$,  the complete spectrum at all KK levels (\ref{eq:KKlevels}) has the following $\textrm{OSp}(4|1) \times \textrm{U}(1)$ representation content. The $\textrm{OSp}(4|1)$ multiplets are those that appear in (\ref{eq:AllSU3Mults_ell0}), for $\ell =0$, and (\ref{eq:AllSU3Mults}), for $\ell \geq 1$, and all $n$ in both cases. The U(1) charges $m$ of these multiplets are obtained by branching the SU(3) representations $[p,q]$ that appear in (\ref{eq:AllSU3Mults_ell0}), (\ref{eq:AllSU3Mults}) under
\begin{equation} \label{eq:SU3ToSO3ToU1}
\textrm{SU}(3) \supset \textrm{SO}(3) \supset \textrm{U}(1) \; , 
\end{equation}
with SO(3) the real subgroup of SU(3) as in (\ref{eq:Sol2Branching}). Thus, these charges are obtained by first branching the $[p,q]$ representation of SU(3) into SO(3) representations of spin $j$ and dimension $(\bm{2 j +1})$ through
\begin{align}	\label{eq: SU3introSO3real}
	[p,q]\rightarrow
	&\bigoplus_{a=0}^{\left[\frac{p}2\right]}\bigoplus_{b=0}^{\left[\frac{q}2\right]}(\bm{2p+2q-4a-4b+1})
	\oplus\,\bigoplus_{a=0}^{\left[\frac{p+1}2\right]-1}\bigoplus_{b=0}^{\left[\frac{q+1}2\right]-1}(\bm{2p+2q-4a-4b-1})\,,	
\end{align}
and then further branching under $\textrm{SO}(3) \supset \textrm{U}(1)$ as usual via
\begin{equation}	\label{eq: SO3intoU1}
	(\bm{2j+1})\rightarrow\bigoplus_{m=-j}^j m\,.
\end{equation}

\begin{figure}
\centering
	\begin{subfigure}{.42\textwidth}
		\centering
		\includegraphics[width=1.0\linewidth]{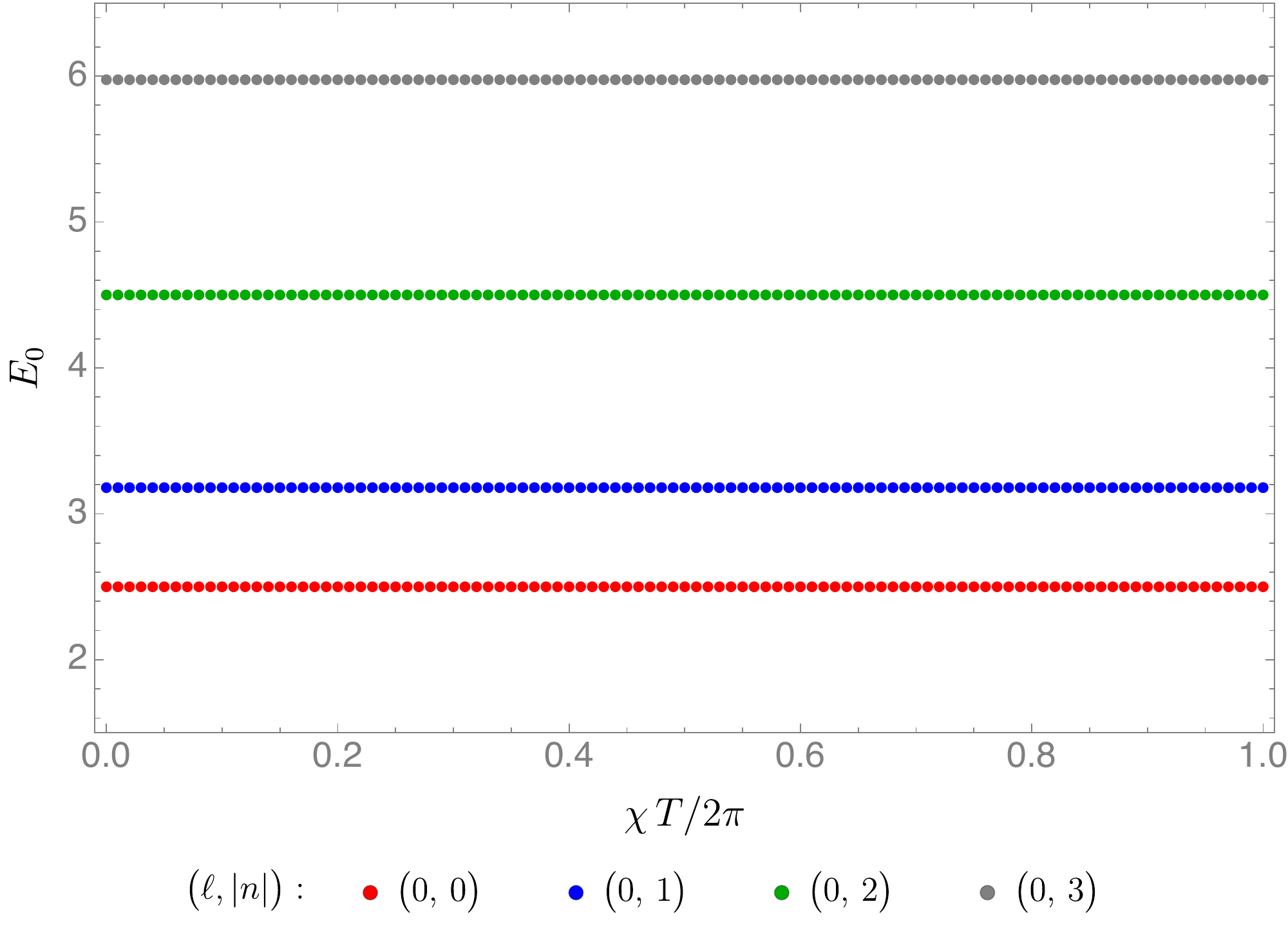}
  		\caption{\footnotesize{(M)GRAVs}}
		\label{fig:SO3gravs}
	\end{subfigure}%
	\quad
	\begin{subfigure}{.42\textwidth}
		\centering
		\includegraphics[width=1.0\linewidth]{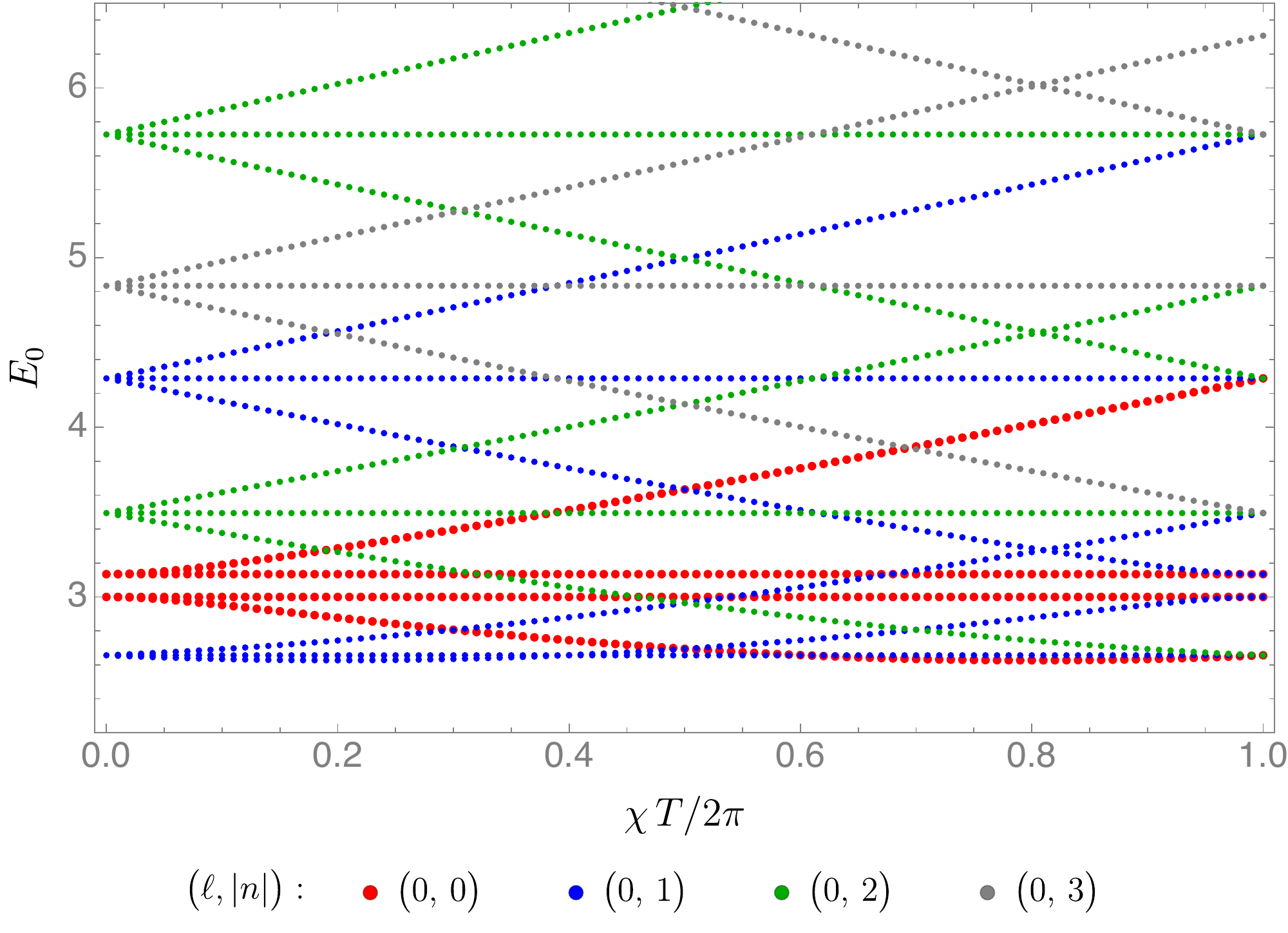}
  		\caption{\footnotesize{GINOs}}
		\label{fig:SO3ginos}
	\end{subfigure}\\[8pt]
	\begin{subfigure}{.42\textwidth}
		\centering
		\includegraphics[width=1.0\linewidth]{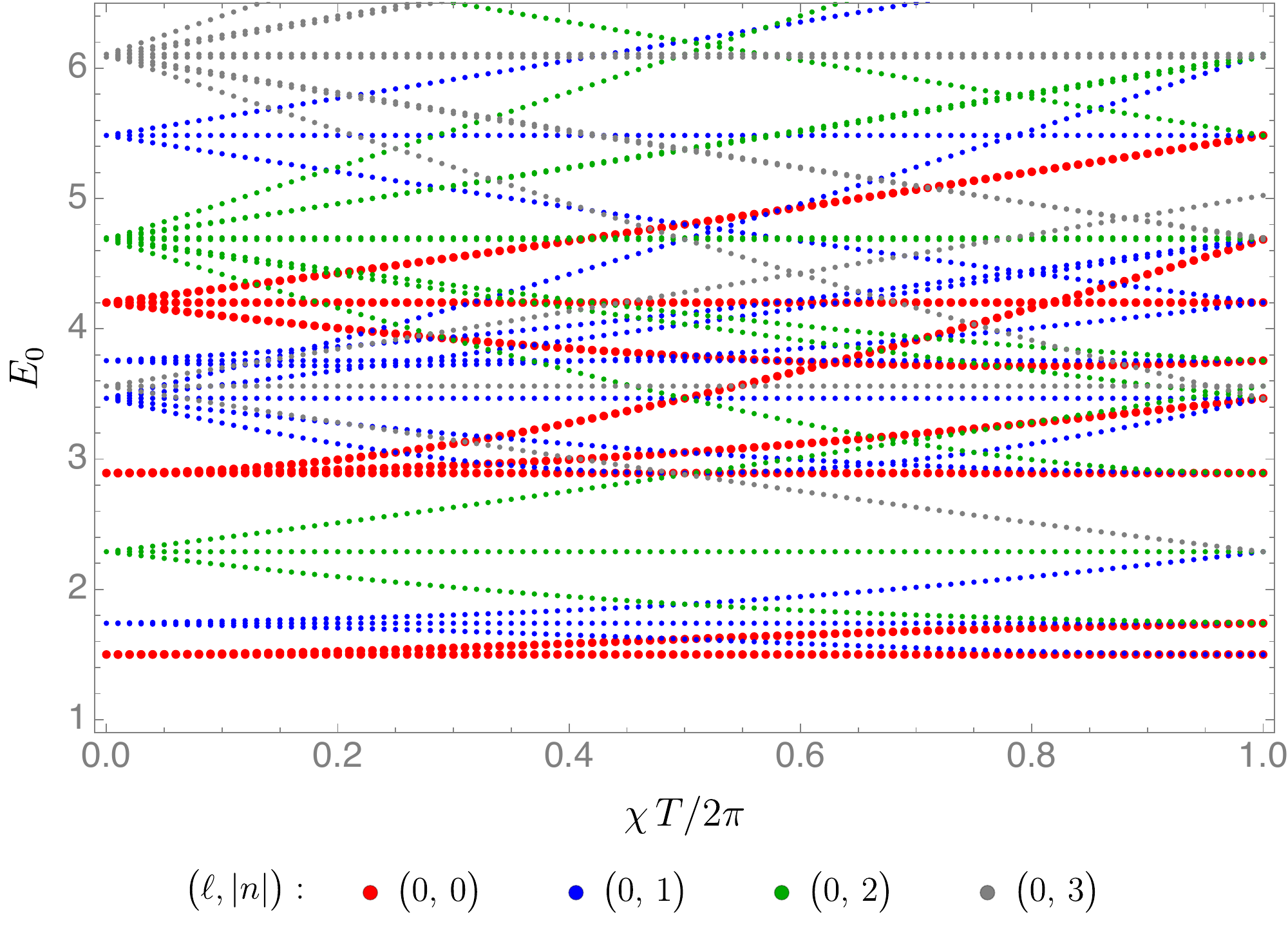}
  		\caption{\footnotesize{(M)VECs}}
		\label{fig:SO3vecs}
	\end{subfigure}%
	\quad
	\begin{subfigure}{.42\textwidth}
		\centering
		\includegraphics[width=1.0\linewidth]{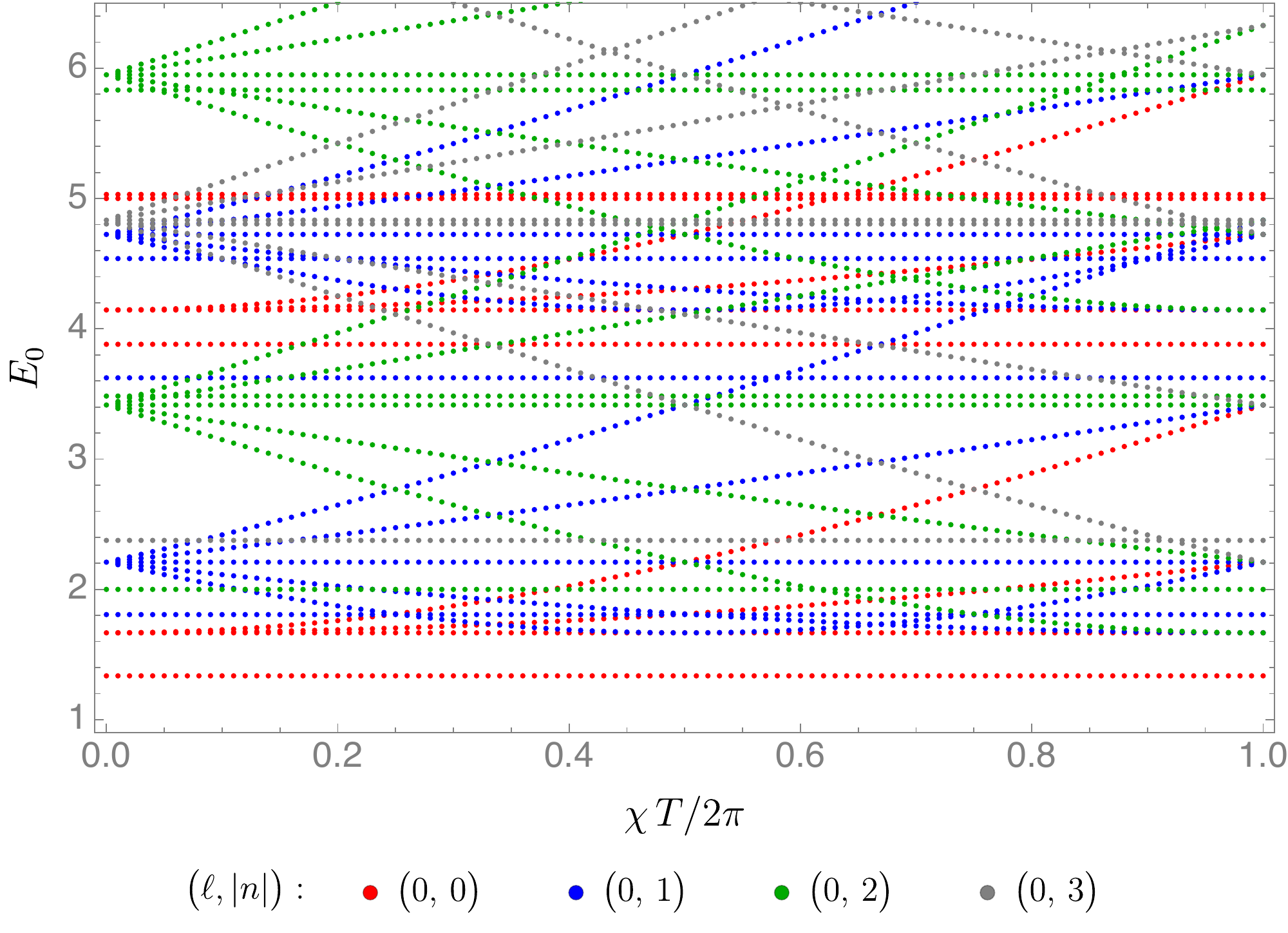}
  		\caption{\footnotesize{CHIRALs}}
		\label{fig:SO3chirals}
	\end{subfigure}
	\caption{Dimensions $E_0$ of all OSp$(4|1)$ multiplets present in the spectrum of the family of solutions of \cite{Berman:2021ynm} (see also \cite{Bobev:2021rtg}) reviewed in section \ref{sec:U(1)toSO3}, at the specified KK levels $(\ell,\, n)$. \label{fig:so3sol} }
\end{figure}

We have obtained numerically the dimensions of these multiplets at a large subset of discretised values of $\chi$ at various KK levels. See figure \ref{fig:so3sol} for a graphical summary of these results, and 
appendix \ref{sec: attachment} for further details.

We have also obtained analytically all the multiplet dimensions at KK levels $\ell=0$ and all $n$. Denoting an OSp$(4|1)$ multiplet MULT with dimension $E_0$ and U(1) charge $m$ that derives from a $\bm{(2 j +1)}$-dimensional representation of SO(3) via (\ref{eq: SO3intoU1}) by 
\begin{equation}	\label{eq: notationU1inSO3}
	{\rm MULT}[E_0;\, m\, (\bm{2j+1})]\, ,
\end{equation}
we have, at $\ell=0$ and all $n$, the following graviton,
\begin{equation} \label{eq:GRAVSO3Fam}
	{\rm \underline{GRAV}}\Big[1+\sqrt{\tfrac94+\tfrac{10\pi^2n^2}{T^2}};\,0\, (\bm{1})\Big]\,,
\end{equation}
and gravitino multiplets,
\begin{align} \label{eq:GINOSO3Fam}
	&{\rm GINO}\Big[1+\tfrac{1}{3 \sqrt{2}}\sqrt{77+45\big(\tfrac{2\pi n}{T}\big)^2+5 \sqrt{1+210\big(\tfrac{2\pi n}{T}\big)^2}};\,0\,(\bm{3})\Big]	\nonumber	\\[5pt]
	&\quad\oplus{\rm GINO}\Big[1+\tfrac{1}{3 \sqrt{2}}\sqrt{77+45\big(\tfrac{2\pi n}{T}\pm\chi\big)^2+5 \sqrt{1+210\big(\tfrac{2\pi n}{T}\pm\chi\big)^2}};\,\pm1\,(\bm{3})\Big]	
	\nonumber\\[5pt]
	&\oplus{\rm GINO}\Big[1+\tfrac{1}{3 \sqrt{2}}\sqrt{77+45\big(\tfrac{2\pi n}{T}\big)^2-5 \sqrt{1+210\big(\tfrac{2\pi n}{T}\big)^2}};\,0\,(\bm{3})\Big]		
	\nonumber\\[5pt]
	&\quad\oplus{\rm GINO}\Big[1+\tfrac{1}{3 \sqrt{2}}\sqrt{77+45\big(\tfrac{2\pi n}{T}\pm\chi\big)^2-5 \sqrt{1+210\big(\tfrac{2\pi n}{T}\pm\chi\big)^2}};\,\pm1\,(\bm{3})\Big]\, .	
\end{align}
Here and elsewhere, the $\pm$ signs in the labels of a given entry are correlated. For example, the second line of (\ref{eq:GINOSO3Fam}) contains two GINO multiplets. The vector multiplets present in the spectrum at $\ell =0$ and all $n$ are
\begin{align} \label{eq:VECSO3Fam}
	&{\rm VEC}\Big[1+\sqrt{\tfrac{43}{12}+\tfrac52\big(\tfrac{2\pi n}{T}\big)^2};\,0\,(\bm{5})\Big]	\nonumber\\[5pt]
	&\quad\oplus{\rm VEC}\Big[1+\sqrt{\tfrac{43}{12}+\tfrac52\big(\tfrac{2\pi n}{T}\pm\chi\big)^2};\,\pm1\,(\bm{5})\Big]	\nonumber\\[5pt]
	&\qquad
	\oplus{\rm VEC}\Big[1+\sqrt{\tfrac{43}{12}+\tfrac52\big(\tfrac{2\pi n}{T}\pm2\chi\big)^2};\,\pm2\,(\bm{5})\Big]	\nonumber\\[5pt]
	&\oplus{\rm \underline{VEC}}\Big[1+\sqrt{\tfrac{1}{4}+\alpha_0};\,0\,(\bm{3})\Big]\oplus{\rm \underline{\underline{VEC}}}\Big[1+\sqrt{\tfrac{1}{4}+\alpha_\pm};\,\pm1\,(\bm{3})\Big]	\nonumber\\[5pt]
	&\quad\oplus{\rm VEC}\Big[1+\sqrt{\tfrac{1}{4}+\beta_0};\,0\,(\bm{3})\Big]\oplus{\rm VEC}\Big[1+\sqrt{\tfrac{1}{4}+\beta_\pm};\,\pm1\,(\bm{3})\Big]	\nonumber\\[5pt]
	&\qquad\oplus{\rm VEC}\Big[1+\sqrt{\tfrac{1}{4}+\gamma_0};\,0\,(\bm{3})\Big]	\oplus{\rm VEC}\Big[1+\sqrt{\tfrac{1}{4}+\gamma_\pm};\,\pm1\,(\bm{3})\Big]\,,
\end{align}
where $\alpha_m,\beta_m,\gamma_m$ ($m = 0 , \pm$), are the roots of the following $m$-, $n$- and $\chi$-dependent (through (\ref{eq:chiComb})) cubic polynomials\footnote{Our \eqref{eq: cubicpoly_vec} agrees with (3.26) of \cite{Bobev:2021rtg} under the identifications $\chi_{\rm here}=\sqrt{\frac{5}{27}}\,\chi_{\rm there}$  and $x_{\rm here}=\alpha_{\rm there}-\frac14$. Despite the fact that $\eta_m$ in \eqref{eq: etavecs} are complex, the expressions for $\alpha_m$, $\beta_m$ and $\gamma_m$ in (\ref{eq: cubicroots_vec}) are always real. This does not seem to be the case for the roots (3.27) in \cite{Bobev:2021rtg}, which appear to have non-zero imaginary parts. The variable $x$ in (\ref{eq:VECSO3Fam}) should not be confused with the embedding tensor component $x$ in (\ref{eq: embtensorcomponents}). \label{ftnt: recoveringBobev}}
\begin{equation}	\label{eq: cubicpoly_vec}
	\begin{aligned}
		\mathcal{P}^{\rm vec}_m(x)&=(x-\alpha_m)(x-\beta_m)(x-\gamma_m)	\\[5pt]
					&=x^3-\Big[\tfrac{15}{2}\big(\tfrac{2\pi n}{T}+m\chi\big)^2+20\Big] x^2+\Big[\tfrac{75}{4}\big(\tfrac{2\pi n}{T}+m\chi\big)^4+\tfrac{950}{27}\big(\tfrac{2\pi n}{T}+m\chi\big)^2+100\Big] x\\
					&\quad-\tfrac{125}{8}\big(\tfrac{2\pi n}{T}+m\chi\big)^6+\tfrac{1000}{27}\big(\tfrac{2\pi n}{T}+m\chi\big)^4-\tfrac{1750}{27}\big(\tfrac{2\pi n}{T}+m\chi\big)^2\, .
	\end{aligned}
\end{equation}
Explicit expressions for these can be found that read
\begin{equation}	\label{eq: cubicroots_vec}
	\begin{aligned}
		\tfrac{9}{5}\alpha_m
		& \equiv  \tfrac{3}{2}\Big[3\big(\tfrac{2\pi n}{T}+m\chi\big)^2+8\Big]-\eta_m^{\nicefrac13}-2 \Big[35\big(\tfrac{2\pi n}{T}+m\chi\big)^2+18\Big]\eta_m^{\nicefrac{-1}3}\,,	\\
		\tfrac{9}{5}\beta_m
		& \equiv \tfrac{3}{2}\Big[3\big(\tfrac{2\pi n}{T}+m\chi\big)^2+8\Big]+e^{i\pi/3}\,\eta_m^{\nicefrac13}+2\,e^{-i\pi/3} \Big[35\big(\tfrac{2\pi n}{T}+m\chi\big)^2+18\Big]\eta_m^{\nicefrac{-1}3}\,,	\\
		\tfrac{9}{5}\gamma_m
		& \equiv  \tfrac{3}{2}\Big[3\big(\tfrac{2\pi n}{T}+m\chi\big)^2+8\Big]+e^{-i\pi/3}\,\eta_m^{\nicefrac13}+2\,e^{i\pi/3} \Big[35\big(\tfrac{2\pi n}{T}+m\chi\big)^2+18\Big]\eta_m^{\nicefrac{-1}3}\,,
	\end{aligned}
\end{equation}
with
\begin{equation}	\label{eq: etavecs}
	\eta_m=216-720\big(\tfrac{2\pi n}{T}+m\chi\big)^2+10i\big(\tfrac{2\pi n}{T}+m\chi\big)\sqrt{5832 + 108 \big(\tfrac{2\pi n}{T}+m\chi\big)^2 + 3430\big(\tfrac{2\pi n}{T}+m\chi\big)^4}\,.
\end{equation}
Finally, the chiral multiplets are
\begin{align}
	&{\rm CHIRAL}\Big[1+\tfrac{1}{3 \sqrt{2}}\sqrt{93+45\big(\tfrac{2\pi n}{T}\big)^2+5\sqrt{17^2+210\big(\tfrac{2\pi n}{T}\big)^2}};\,0\,(\bm{5})\Big]	\nonumber\\[5pt]
	&\quad\oplus{\rm CHIRAL}\Big[1+\tfrac{1}{3 \sqrt{2}}\sqrt{93+45\big(\tfrac{2\pi n}{T}\pm\chi\big)^2+5\sqrt{17^2+210\big(\tfrac{2\pi n}{T}\pm\chi\big)^2}};\,\pm1\,(\bm{5})\Big]	\nonumber\\[5pt]
	&\qquad\oplus{\rm CHIRAL}\Big[1+\tfrac{1}{3 \sqrt{2}}\sqrt{93+45\big(\tfrac{2\pi n}{T}\pm2\chi\big)^2+5\sqrt{17^2+210\big(\tfrac{2\pi n}{T}\pm2\chi\big)^2}};\,\pm2\,(\bm{5})\Big]	\nonumber\\[5pt]
	&\oplus{\rm CHIRAL}\Big[1+\tfrac{1}{3 \sqrt{2}}\sqrt{93+45\big(\tfrac{2\pi n}{T}\big)^2-5\sqrt{17^2+210\big(\tfrac{2\pi n}{T}\big)^2}};\,0\,(\bm{5})\Big]	\nonumber\\[5pt]
	&\quad\oplus{\rm CHIRAL}\Big[1+\tfrac{1}{3 \sqrt{2}}\sqrt{93+45\big(\tfrac{2\pi n}{T}\pm\chi\big)^2-5\sqrt{17^2+210\big(\tfrac{2\pi n}{T}\pm\chi\big)^2}};\,\pm1\,(\bm{5})\Big]	\nonumber\\[5pt]
	&\qquad\oplus{\rm CHIRAL}\Big[1+\tfrac{1}{3 \sqrt{2}}\sqrt{93+45\big(\tfrac{2\pi n}{T}\pm2\chi\big)^2-5\sqrt{17^2+210\big(\tfrac{2\pi n}{T}\pm2\chi\big)^2}};\,\pm2\,(\bm{5})\Big]	\nonumber\\[5pt]
	&\oplus\bigoplus_{i=1}^4{\rm CHIRAL}\Big[1+\sqrt{\omega_{i}};\,0\,(\bm{1})\Big]\,,
\end{align}
where $\omega_i$, $i=1,2,3,4$, are now the roots of the $n$-dependent (but now $\chi$-independent, consistent with (\ref{eq:chiComb})) quartic polynomials
\begin{equation}	\label{eq: quarticpoly_chiral}
	\begin{aligned}
		\mathcal{P}^{\rm chiral}(x)&=(x-\omega_1)(x-\omega_2)(x-\omega_3)(x-\omega_4)	\\[5pt]
					&=x^4-\Big[10\big(\tfrac{2\pi n}{T}\big)^2+\tfrac{122}{3}\Big] x^3
					+\Big[\tfrac{75}{2}\big(\tfrac{2\pi n}{T}\big)^4+\tfrac{3860}{27}\big(\tfrac{2\pi n}{T}\big)^2+\tfrac{43121}{81}\Big] x^2\\[5pt]
					&\quad-\Big[\tfrac{125}{2}\big(\tfrac{2\pi n}{T}\big)^6-\tfrac{2575}{54}\big(\tfrac{2\pi n}{T}\big)^4+\tfrac{22910}{27}\big(\tfrac{2\pi n}{T}\big)^2+\tfrac{179684}{81}\Big] x	\\[5pt]
					&\qquad+\tfrac{625}{16}\big(\tfrac{2\pi n}{T}\big)^8-\tfrac{20375}{54}\big(\tfrac{2\pi n}{T}\big)^6+\tfrac{93725}{81}\big(\tfrac{2\pi n}{T}\big)^4+\tfrac{52960}{81}\big(\tfrac{2\pi n}{T}\big)^2+\tfrac{6592}{27}
					\,.
	\end{aligned}
\end{equation}
At lowest level, $n=0$, this quartic polynomial reduces to that reported in \cite{Berman:2021ynm,Bobev:2021rtg},
\begin{equation}	\label{eq: quarticpoly_zero}
	\tfrac1{81}(x-16)(81x^3-1998x^2+11153x-1236)\,.
\end{equation}

As our numerics suggest and the above analytic results confirm for the reported KK levels, the dependence of the multiplet dimensions with the KK level $n$, the modulus $\chi$, and the U(1) charge $m$ are locked into the combination (\ref{eq:chiComb}). As discussed generally in section \ref{sec:CommonAspects}, the underlined GRAV multiplet (\ref{eq:GRAVSO3Fam}) becomes massless at $n=0$ for all $\chi$ via (\ref{eq: shortening_generalGRAV}). Similarly, the singly underlined VEC in (\ref{eq:VECSO3Fam}) becomes massless at $n=0$ for all $\chi$ by (\ref{eq: shortening_generalVEC}), further producing a CHIRAL multiplet that contains the modulus $\chi$. At the SO(3) symmetry enhanced points, the doubly underlined VEC multiplets in (\ref{eq:VECSO3Fam}) further become massless and together with the former furnish the adjoint of SO(3). This happens either at KK level $n=0$ or at different KK levels $n^\prime$ as in (\ref{eq:KKreshuffle}), depending on whether the SO(3) point is attained at $\chi=0$ or at a location periodically identified with the former through (\ref{eq:periodicitySO3Fam}). Except for the disagreement in the VEC multiplet dimensions noted in footnote \ref{ftnt: recoveringBobev}, our $\ell=n=0$ spectrum recovers that reported in \cite{Berman:2021ynm,Bobev:2021rtg}. 

We have also determined analytically the GRAV multiplet dimensions present at KK levels $\ell=1$ and all $n$. We have 
\begin{equation}
	\begin{aligned}
	&{\rm GRAV}\Big[1+\sqrt{\tfrac{271}{36}+\tfrac5{2}\big(\tfrac{2\pi n}{T}\big)^2-\tfrac{10}{9}\sqrt{4+3\big(\tfrac{2\pi n}{T}\big)^2}};\, 0\, (\bm{3})\Big]		\\
	&\quad\oplus
	{\rm GRAV}\Big[1+\sqrt{\tfrac{271}{36}+\tfrac5{2}\big(\tfrac{2\pi n}{T}\pm\chi\big)^2-\tfrac{10}{9}\sqrt{4+3\big(\tfrac{2\pi n}{T}\pm\chi\big)^2}};\, \pm1\, (\bm{3})\Big]		\\
	&\oplus
	{\rm GRAV}\Big[1+\sqrt{\tfrac{271}{36}+\tfrac5{2}\big(\tfrac{2\pi n}{T}\big)^2+\tfrac{10}{9}\sqrt{4+3\big(\tfrac{2\pi n}{T}\big)^2}};\, 0\, (\bm{3})\Big]		\\
	&\quad\oplus
	{\rm GRAV}\Big[1+\sqrt{\tfrac{271}{36}+\tfrac5{2}\big(\tfrac{2\pi n}{T}\pm\chi\big)^2+\tfrac{10}{9}\sqrt{4+3\big(\tfrac{2\pi n}{T}\pm\chi\big)^2}};\, \pm1\, (\bm{3})\Big]\,.
	\end{aligned}
\end{equation}
%

\subsection{One-parameter $\textrm{U}(1)$-invariant solution with no enhancement}	\label{sec:U(1)KK}

We conclude this section with the discussion of the spectrum on the one-parameter family of U(1)-invariant solutions of \cite{Bobev:2021rtg}, reviewed in section \ref{sec:U(1)} above. The content of $\textrm{OSp}(4|1) \times \textrm{U}(1)$ multiplets present in this spectrum is the same outlined at the beginning of section \ref{sec:U(1)toSO3KK}, because both families of solutions share the same residual U(1) symmetry.  Namely, the complete KK spectrum at all levels (\ref{eq:KKlevels}) is given by (\ref{eq:AllSU3Mults_ell0}), (\ref{eq:AllSU3Mults}), with the SU(3) representations therein branched out into U(1) representations under (\ref{eq:SU3ToSO3ToU1}) via (\ref{eq: SU3introSO3real}), (\ref{eq: SO3intoU1}). The $\ell=0$ GRAV multiplet 
in (\ref{eq:AllSU3Mults_ell0}) becomes massless at $n=0$ for all $\chi$ through (\ref{eq: shortening_generalGRAV}). Similarly, the U(1)-singlet arising from the underlined $\ell=0$ VEC multiplet in (\ref{eq:AllSU3Mults_ell0}) also splits at $n=0$ and for all $\chi$ into an MVEC and a CHIRAL multiplet via (\ref{eq: shortening_generalVEC}). As usual, these massless multiplets respectively contain the graviton, the gauge field and the modulus $\chi$. Unlike for the previous solutions, in this case there are no symmetry enhancements across the one-dimensional moduli space.

\begin{figure}
\centering
	\begin{subfigure}{.42\textwidth}
		\centering
		\includegraphics[width=1.0\linewidth]{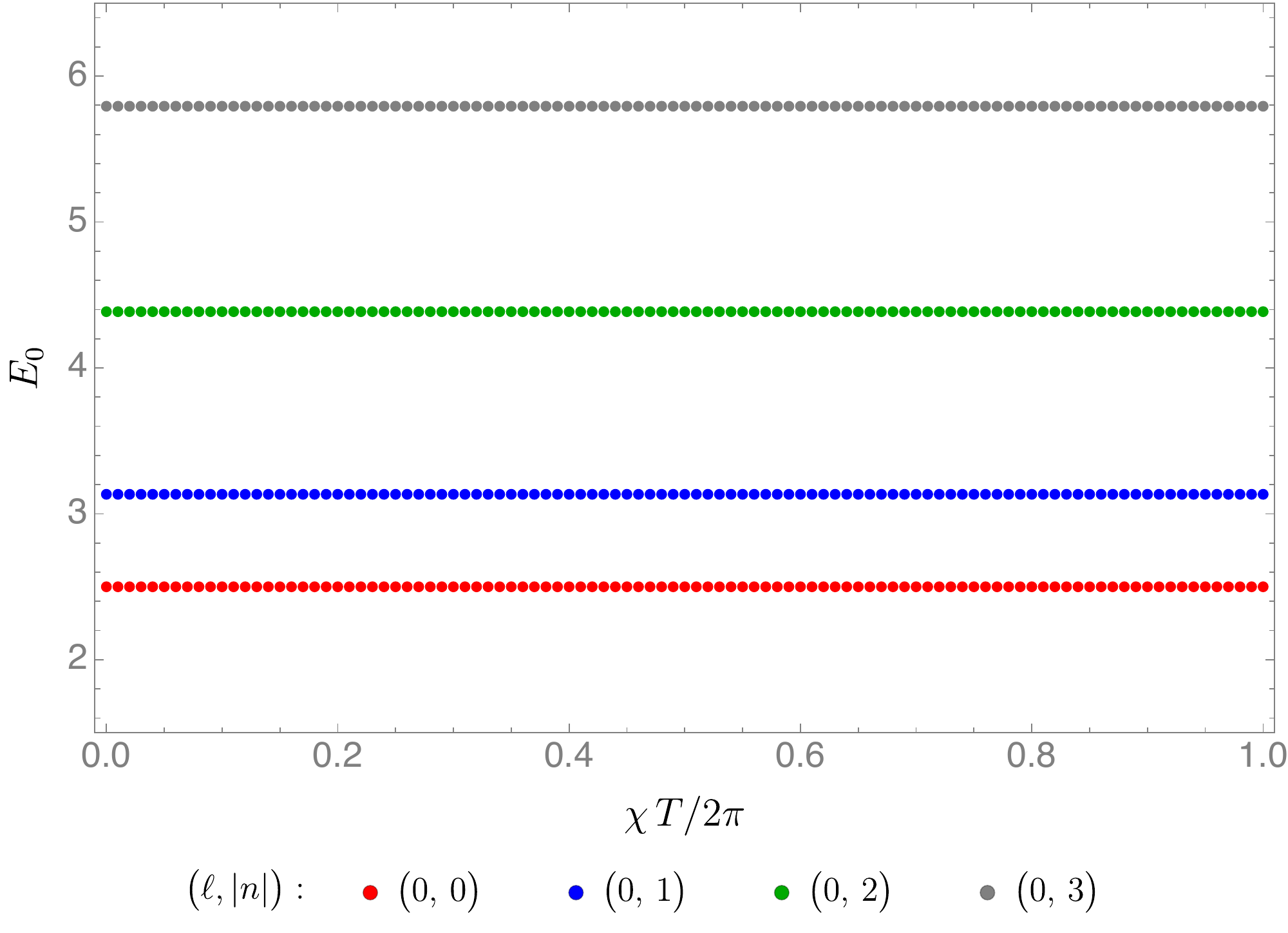}
  		\caption{\footnotesize{(M)GRAVs}}
		\label{fig:U1gravs}
	\end{subfigure}%
	\quad
	\begin{subfigure}{.42\textwidth}
		\centering
		\includegraphics[width=1.0\linewidth]{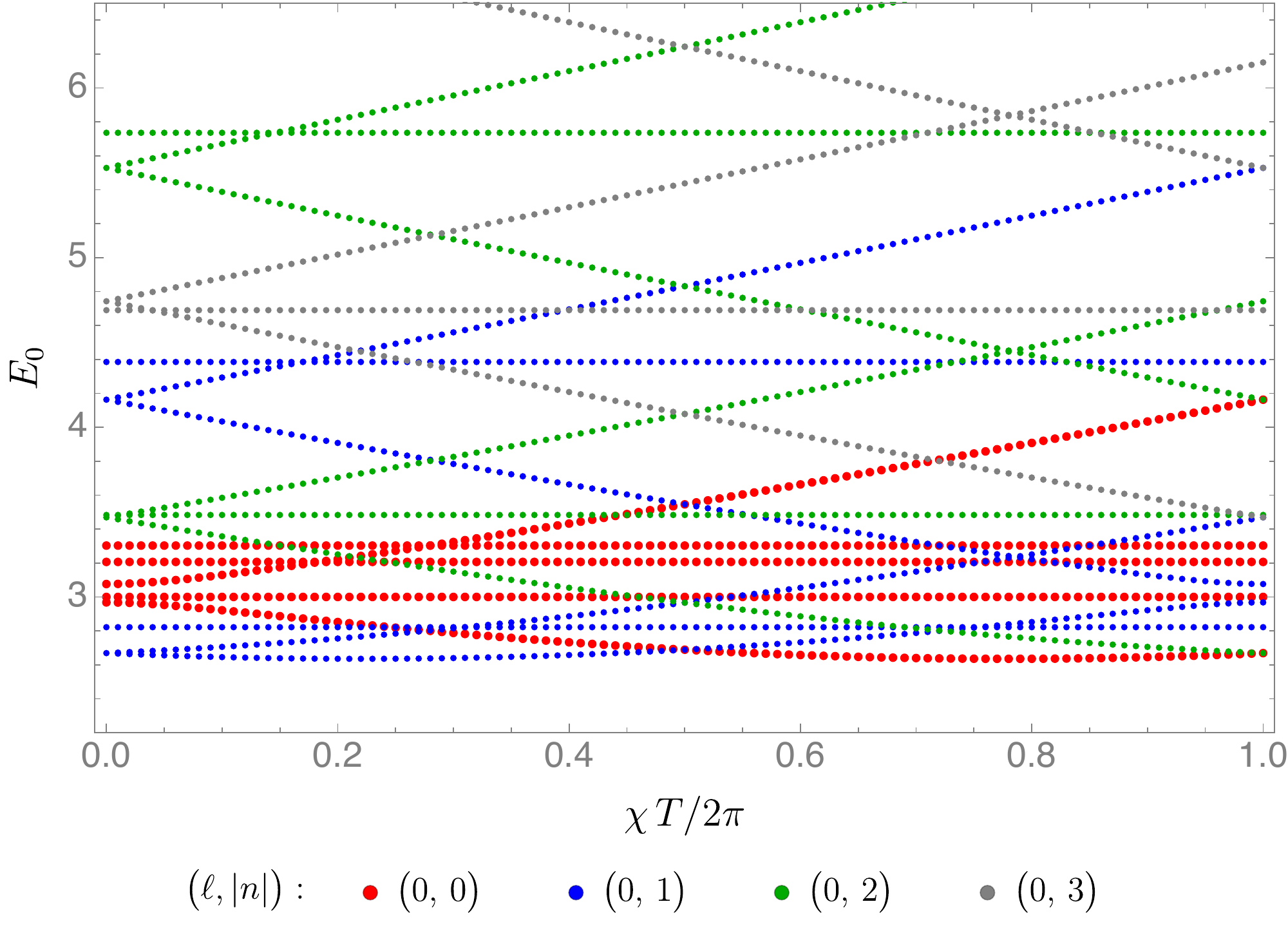}
  		\caption{\footnotesize{GINOs}}
		\label{fig:U1ginos}
	\end{subfigure}\\[8pt]
	\begin{subfigure}{.42\textwidth}
		\centering
		\includegraphics[width=1.0\linewidth]{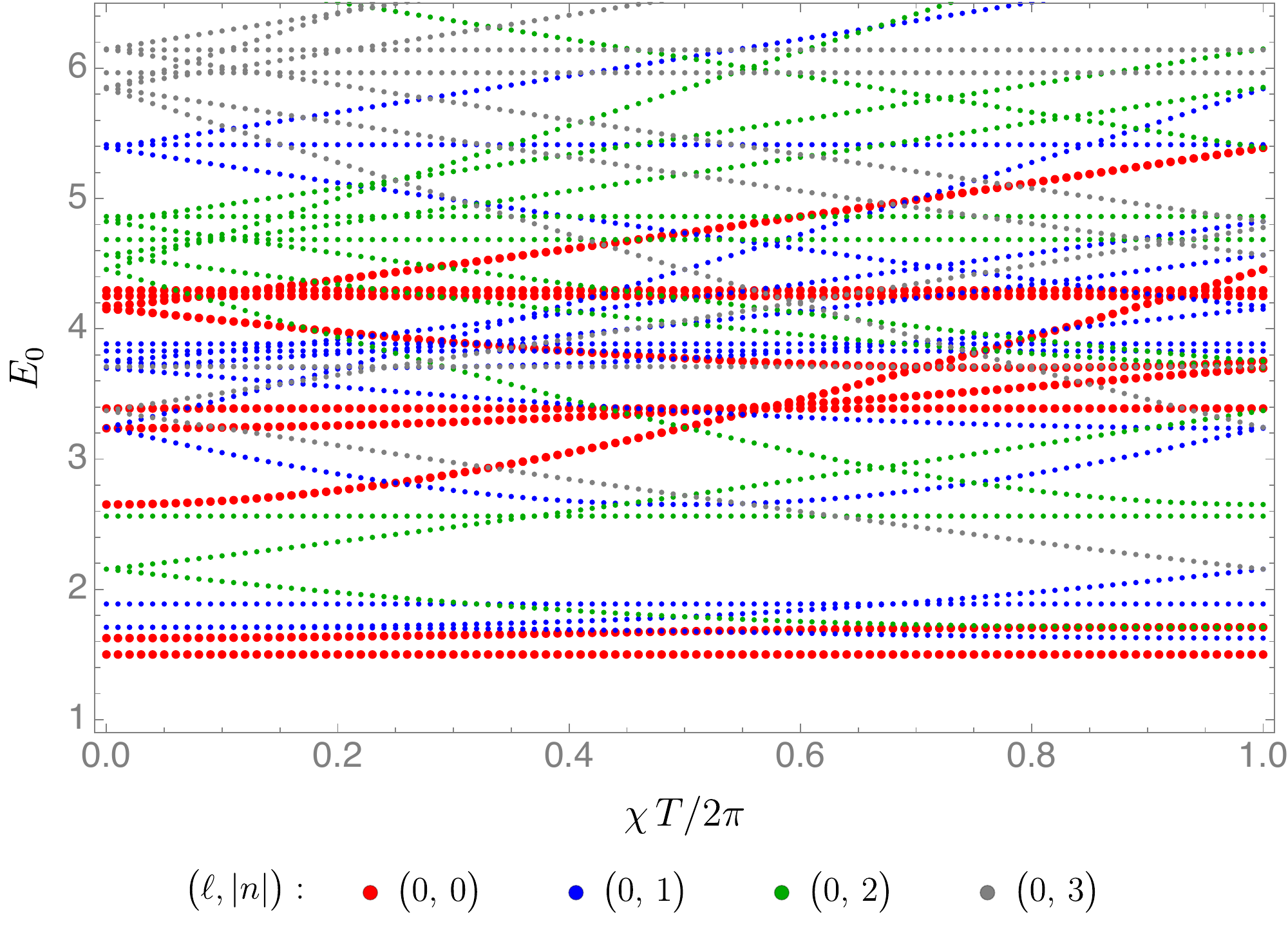}
  		\caption{\footnotesize{(M)VECs}}
		\label{fig:U1vecs}
	\end{subfigure}%
	\quad
	\begin{subfigure}{.42\textwidth}
		\centering
		\includegraphics[width=1.0\linewidth]{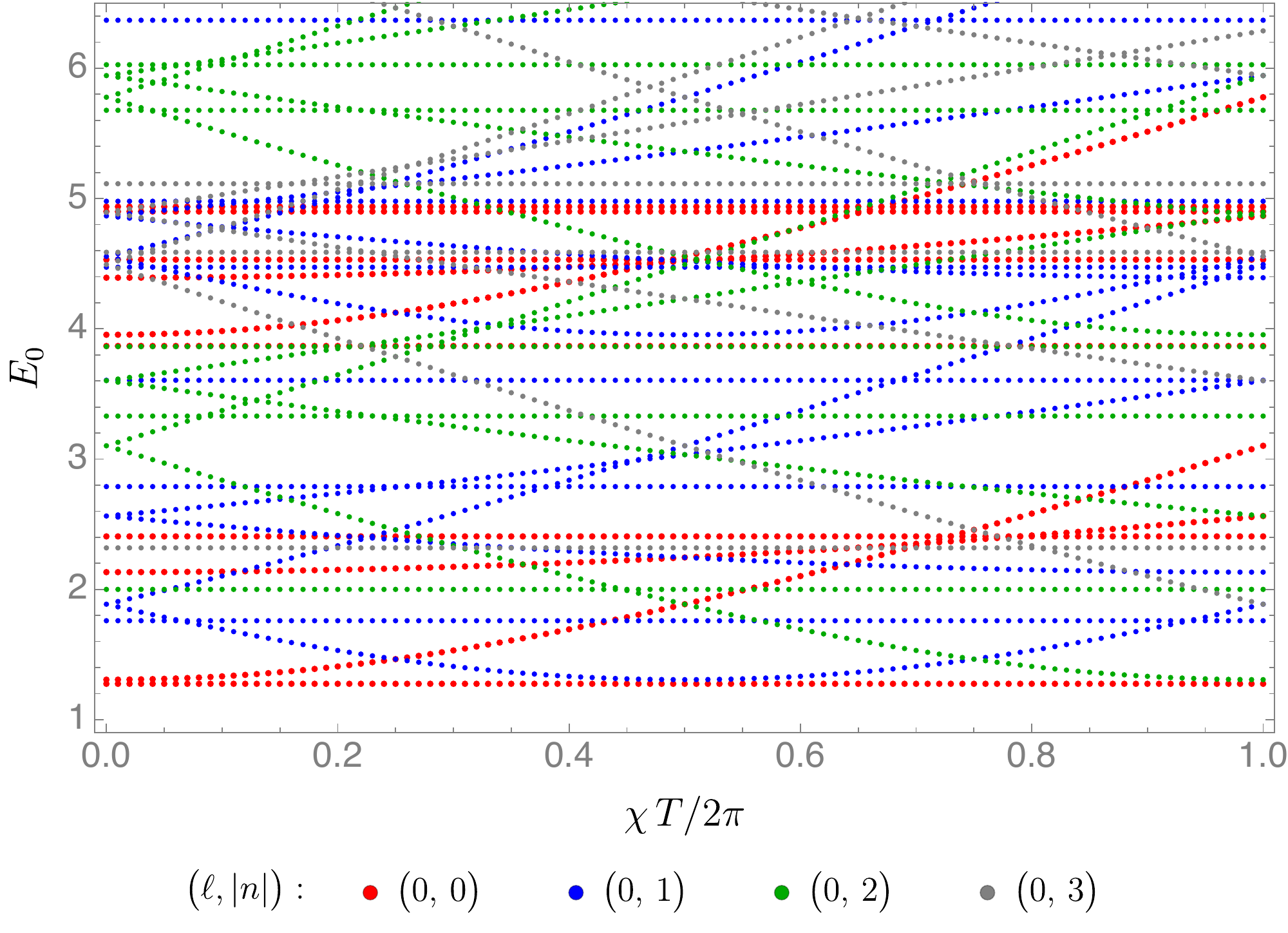}
  		\caption{\footnotesize{CHIRALs}}
		\label{fig:U1chirals}
	\end{subfigure}
	\caption{Dimensions $E_0$ of all OSp$(4|1)$ multiplets present in the spectrum of the family of solutions of \cite{Bobev:2021rtg} reviewed in section \ref{sec:U(1)}, at the specified KK levels $(\ell,\, n)$.\label{fig:U1NoEnhancement} }
\end{figure}

Of course, while the multiplet structure is the same, the multiplet dimensions are different for this family than those reported in section \ref{sec:U(1)toSO3KK} for the family reviewed in section \ref{sec:U(1)toSO3}. Most of our results for the supermultiplet dimensions for this family of solutions at various KK levels are numerical. See figure \ref{fig:U1NoEnhancement} for a graphical account and appendix \ref{sec: attachment} for further details. Our numerical results endorse also for this family the generic periodic behaviour in $\chi$ discussed in section \ref{sec:CommonAspects}. 

We do have some analytic results for this family as well. More concretely, we have obtained analytically  the GRAV multiplet dimensions at KK levels $\ell=0$ and $\ell=1$, for all $n$ in both cases. Denoting an OSp$(4|1)$ multiplet MULT with dimension $E_0$ and U(1) charge $m$ by 
\begin{equation}	\label{eq: notationU1noenhancement}
	{\rm MULT}[E_0;\, m]\, ,
\end{equation}
the GRAV multiplet present in the spectrum at $\ell=0$ and all $n$ is
\begin{equation}
	{\rm \underline{GRAV}}\Big[1+\sqrt{\tfrac94+\tfrac{54}{\left(\sqrt{13}-2\right) \sqrt{134+22\sqrt{13}}}\big(\tfrac{2\pi n}{T}\big)^2};\, 0\Big]\,.
\end{equation}
At $\ell=1$ and all $n$ we have
\begin{equation} \label{eq:GRAVsell=1Alln}
	\begin{aligned}
	&{\rm GRAV}\Big[1+\sqrt{\tfrac94+\tfrac{\sqrt{23+103 \sqrt{13}}\,+\sqrt{24 \big(\tfrac{2\pi n}{T}\big)^2+23+7 \sqrt{13}}\,+2 \sqrt{3 \sqrt{13}+3} \big(\tfrac{2\pi n}{T}\big)^2}{2 \sqrt{\sqrt{13}-1}}};\, 0\Big]		\\
	& \oplus
	{\rm GRAV}\Big[1+\sqrt{\tfrac94+\tfrac{\sqrt{23+103 \sqrt{13}}\,-\sqrt{24 \big(\tfrac{2\pi n}{T}\big)^2+23+7 \sqrt{13}}\,+2 \sqrt{3 \left(\sqrt{13}+1\right)} \big(\tfrac{2\pi n}{T}\big)^2}{2 \sqrt{\sqrt{13}-1}}};\, 0\Big]		\\
	&\oplus
	{\rm GRAV}\Big[1+\sqrt{\tfrac94+\tfrac{\sqrt{2558 \sqrt{13}-353}\,+\sqrt{144 \left(\sqrt{13}+5\right) \big(\tfrac{2\pi n}{T}\pm\chi\big)^2+49 \left(5 \sqrt{13}+19\right)}\,+8 \sqrt{6 \left(\sqrt{13}+1\right)} \big(\tfrac{2\pi n}{T}\pm\chi\big)^2}{8 \sqrt{2}\sqrt{\sqrt{13}-1}}};\, \pm1\Big]			\\
	& \oplus
	{\rm GRAV}\Big[1+\sqrt{\tfrac94+\tfrac{\sqrt{2558 \sqrt{13}-353}\,-\sqrt{144 \left(\sqrt{13}+5\right) \big(\tfrac{2\pi n}{T}\pm\chi\big)^2+49 \left(5 \sqrt{13}+19\right)}\,+8 \sqrt{6 \left(\sqrt{13}+1\right)} \big(\tfrac{2\pi n}{T}\pm\chi\big)^2}{8 \sqrt{2}\sqrt{\sqrt{13}-1}}};\, \pm1\Big]\,.
	\end{aligned}
\end{equation}

\newpage


\section{Discussion} \label{sec:Discussion}


We have assessed the KK spectrum of three specific families of $\cN=1$ S-fold solutions (\ref{eq:AdS4ixS5xS1}) of type IIB supergravity that uplift consistently from the AdS vacua of $D=4$ $\cN=8$ supergravity with gaugings (\ref{eq:GaugeGroups}) found in \cite{Guarino:2019oct,Guarino:2020gfe,Berman:2021ynm,Bobev:2021rtg}. For all three cases, we have provided the complete algebraic structure of supermultiplets at all KK levels, and also their dimensions in some cases. Remarkably enough, for the U$(1)^2$-invariant family of \cite{Guarino:2019oct,Guarino:2020gfe}, we have managed to also give analytically the expression, (\ref{eq:GenericE0}), for the supermultiplet dimensions at all KK levels. It is interesting to note that this expression conforms to a generic formula, (1.1) of \cite{Cesaro:2020soq}, known to hold for the supermultiplet dimensions of other similar spectra. For the U(1)-invariant one-parameter families in the $G_+$ gauging \cite{Berman:2021ynm,Bobev:2021rtg}, we have computed the dimensions numerically at discretised values of the modulus that labels the families and at the first few KK levels, see appendix \ref{sec: attachment}. For these two families, we have obtained analytically some supermultiplet dimensions as well. The latter dimensions tend to arise as roots of polynomials of degree higher than two and, for this reason, the resulting expressions (\ref{eq:GRAVSO3Fam})--(\ref{eq:GRAVsell=1Alln}) do not typically conform to the generic formula of \cite{Cesaro:2020soq}. It would be interesting to understand why that formula \cite{Cesaro:2020soq} works in some cases, like the solutions of \cite{Guarino:2019oct,Guarino:2020gfe}, but not in others, \cite{Berman:2021ynm,Bobev:2021rtg}. See \cite{Cesaro:2021haf} for a related discussion.

We also note the following curious fact for the spectrum of the U(1)-invariant cases at hand \cite{Berman:2021ynm,Bobev:2021rtg}. For these spectra, there are many individual KK states of spin-$1/2$ whose masses (and dimensions) differ by one. Some of these states pair up into VEC multiplets, together with appropriate KK vector and scalar states (see {\it e.g.}~table 1 of \cite{Cesaro:2020soq} for the VEC multiplet content). But not all of them do, and the excess goes into different CHIRAL multiplets whose (superconformal primary) dimensions differ by one. There is a protection mechanism (supersymmetry) for this pattern of dimensions of the spin-$1/2$ states that belong to the same VEC multiplet. In contrast, there is no obvious mechanism leading to such type of CHIRALs in the spectrum. The number of CHIRALs with dimensions differing by one grows with the KK level $\ell$, but not all the CHIRALs at fixed level $\ell$ are of this type. We are not aware of the presence of this type of CHIRAL multiplets in the KK spectra \cite{Cesaro:2020soq,Cesaro:2021haf} of other $\cN=1$ AdS$_4$ solutions of M-theory or type IIA that uplift from $D=4$ $\cN=8$ supergravity.

Our results for the KK spectra under consideration confirm that the uplifts of these families of solutions follow the pattern of similar solutions in this class \cite{Giambrone:2021zvp,Guarino:2021kyp,Cesaro:2021tna}, in that the pseudoscalar deformations produce a local fibration (i.e. a Wilson line, see \cite{Berman:2021ynm}) of the $S^5$ in (\ref{eq:AdS4ixS5xS1}) over the $S^1$. The dependence of the masses on the $n$ quantum number and the deformations is locked into the combinations in \eqref{eq:chiComb}. Interestingly, for the geometries at hand, this observation allows one to work backwards and fully reconstruct the type IIB solutions from their (graviton) spectra. Let us sketch the argument, leaving an explicit application of these `reverse engineering' methods for appendix \ref{sec:UpliftSU3Fam}. 

The KK graviton masses are the eigenvalues of the following second-order differential operator \cite{Bachas:2011xa}
\begin{equation}	\label{eq:BachasEstes}
	\mathcal{L}=-\frac{e^{-8A}}{\sqrt{\bar{g}_6}}\partial_m\Big[e^{8A}\sqrt{\bar{g}_6}\,\bar{g}^{mn}\partial_n\Big]
\end{equation}
associated to the warped ten-dimensional metric $ds^2_{10}=e^{2A}(ds^2({\rm AdS}_4)+d\bar{s}^2_6)$ in the Einstein frame. The dependence \eqref{eq:chiComb} implies that the only modification on the operator (\ref{eq:BachasEstes}) corresponding to the deformed solutions with respect to the undeformed solutions with vanishing moduli follows from the replacement
\begin{equation} \label{eq:DerRep}
	\partial_\eta\to\partial_\eta+\tfrac12(\chi_1-\chi_2)k_3+\tfrac12(\chi_1+\chi_2)k_8\,,	
	\qquad\text{or}\qquad
	\partial_\eta\to\partial_\eta+\chi\,\partial_\beta\,.
\end{equation}
Here, $k_3=2\,\partial_\phi$ and $k_8=2(3\,\partial_\psi-\partial_\tau)$ are the Killing vectors corresponding to the Cartan subalgebra of SU(3) in (a suitable rescaling of) the usual Gell-Mann basis, see {\it e.g.}~(A.1) in \cite{Larios:2019kbw} for the embedding of these generators in SO(6). These Killing vectors are normalised so that their eigenvalues are given by the U$(1)^2$ charges \eqref{eq:SU(3)toCartans}.
In (\ref{eq:DerRep}), $\phi$, $\psi$, $\tau$ and $\beta$ are angles on $S^5$, see appendix \ref{sec:UpliftSU3Fam}. From \eqref{eq:BachasEstes}, these deformations can be traced back to the metric and simply amount to the substitutions
\begin{equation}	\label{eq: deformedSU3}
	\phi \to \phi-(\chi_1-\chi_2)\,\eta\,,	\quad
	\psi \to \psi-3(\chi_1+\chi_2)\,\eta\,,	\quad
	\tau \to \tau+(\chi_1+\chi_2)\,\eta
\end{equation}
for the type IIB uplift of the family in section \ref{sec:U(1)2toSU3}, and
\begin{equation} \label{eq:shiftbeta}
	\beta \to \beta-\chi \,\eta\,,
\end{equation}
for the uplifts of the other two families. Solutions of the full type IIB supergravity can be then obtained by extending these replacements to the supergravity forms.

\section*{Acknowledgements}

We would like to thank Alfredo Giambrone for useful comments and discussions. GL would like to thank the Leinweber Center for Theoretical Physics, Ann Arbor, Michigan, for hospitality. The numerical computations of this paper were performed on the Hydra cluster of IFT.  MC is supported by a La Caixa Foundation (ID100010434) predoctoral fellowship LCF/ BQ/DI19/11730027. GL is supported by an FPI-UAM predoctoral fellowship, and was partially supported by the Spain-US Fulbright Commission during the early stages of this work. OV is supported by the NSF grant PHY-2014163. All of us are partially supported through the grants CEX2020-001007-S and PGC2018-095976-B-C21, funded by MCIN/AEI/10.13039/501100011033 and by ERDF A way of making Europe, and through the grant RYC-2015-18741 funded by MCIN/AEI/10.13039/501100011033 and by ESF Investing in your future.



\appendix

\newpage


\section{Ancillary files: numerical spectra}	\label{sec: attachment}


This article comes with three companion Wolfram Mathematica files,
\begin{center}
\texttt{KKSpectrum.nb} \,, \quad
\texttt{KKSpectrum$\_$FamilySO3.wl}  \,,  \quad
\texttt{KKSpectrum$\_$FamilyU1.wl}  \,,
\end{center}
which provide our numerical results for the KK spectra discussed in sections \ref{sec:U(1)toSO3KK} and \ref{sec:U(1)KK} for the one-parameter families of solutions \cite{Berman:2021ynm,Bobev:2021rtg} reviewed in sections \ref{sec:U(1)toSO3} and \ref{sec:U(1)}, respectively. The first of these files, with \texttt{nb} extension, provides a user interface, while the last two \texttt{wl} files contain the proper databases which can be accessed with the former. All three files must be downloaded into the same local folder before the \texttt{nb} file can be executed. The data contained in the \texttt{wl} files must be loaded into memory by running the \texttt{Get} commands in the \texttt{nb} file. The `\texttt{SO3}'  label in the first \texttt{wl} file (and in the corresponding function below) refers to the fact that the family of section \ref{sec:U(1)toSO3} contains a point with SO(3)-enhanced symmetry, even though the generic symmetry of this family is only U(1). 

The databases can be accessed by executing from the \texttt{nb} file either of the following two functions, with syntax:
\begin{center}
\texttt{KKSpecSO3}$[\chi , \ell , n ]$ , \qquad
\texttt{KKSpecU1}$[\chi , \ell , n ]$ .
\end{center}
These respectively provide the spectrum of OSp$(4|1)$ multiplets at the value $\chi$ of the parameter that labels the families of sections \ref{sec:U(1)toSO3} and \ref{sec:U(1)}, at $S^5$ and $S^1$ KK levels $\ell$ and $n$. The KK levels must be contained in $\ell\in\{0,1,2,3\}$ and $n\in\{0,\pm1,\pm2,\pm3\}$. 
For both functions, the argument $\chi$ can be any number contained in $0\leq \chi\leq1$, in steps of $\Delta\chi=0.01$. We fix the internal $S^1$ radius to $T=2\pi$ for our numerics.

The output of both functions is the list of OSp$(4|1)$ multiplets present in the KK spectrum at the specified KK levels $\ell$, $n$ and location $\chi$ on the corresponding family of solutions. More concretely, the functions provide the eigenvalues of the KK mass matrices at the requested KK levels on the selected point $\chi$, translated into conformal dimensions, and repacked into supermultiplets of OSp$(4|1)$. The functions do not keep track of the U(1) (or SO(3)) charges of these multiplets, and only tally up their multiplicities. For this reason, the output OSp$(4|1)$ multiplets are simply displayed as
\begin{center}
\texttt{MULT[$\mathtt{E_0}$,deg]} , 
\end{center}
Here, \texttt{MULT} is one of the acronyms for the OSp$(4|1)$ multiplets in the conventions of \cite{Cesaro:2020soq}, which we also use in the main text, $\mathtt{E_0}$ is the dimension and \texttt{deg} the multiplicity. If the input arguments do not meet the above specifications, an error message is printed. 

Our calculations were performed with Mathematica's default machine precision of fifteen decimal places. For simplicity of presentation, the database contained in the \texttt{wl} file is truncated to six digits.

\newpage


\section{Further details and checks on the spectra}	\label{sec: lowKKSU3}


Equations \eqref{eq:AllSU3Mults_ell0} and \eqref{eq:AllSU3Mults} of the main text give the putative representations of (\ref{eq:Fullsusy}) from which the algebraic content of all spectra discussed in this paper follows. For bookkeeping purposes, it is convenient to encode the content of those equations in the following fugacities:
\begin{equation}	\label{eq: SU3fugacities}
	\begin{aligned}
		\nu_{(\nicefrac32)}	&=\frac{y^{\ell+1}-x^{\ell+1}}{y-x}\,,		\\
		\nu_{\1}&=(x+y)\frac{y^{\ell+1}-x^{\ell+1}}{y-x}+(x+y+2)\frac{y^{\ell}-x^{\ell}}{y-x}\,,		\\
		\nu_{(\nicefrac12)}	&= (xy+x+y+1-\delta_{\ell,0})\frac{y^{\ell+1}-x^{\ell+1}}{y-x}		\\
				&\quad+(x^2+y^2+x+y+2)\frac{y^{\ell}-x^{\ell}}{y-x}+(xy+x+y+1)H(\ell)\frac{y^{\ell-1}-x^{\ell-1}}{y-x}\,,		\\
		\nu_{\0}&= (x^2+y^2+2)\frac{y^{\ell+1}-x^{\ell+1}}{y-x}		\\
				&\quad+(2xy+x+y)\frac{y^{\ell}-x^{\ell}}{y-x}+(x^2+y^2+x+y+2)H(\ell)\frac{y^{\ell-1}-x^{\ell-1}}{y-x}\,,		\\
	\end{aligned}
\end{equation}
where $H(\ell)$ is the Heaviside step function. The subindices on each fugacity $\nu$ refer to the spin $s_0$ of the superconformal primary of each type of OSp$(4|1)$ multiplet: from top to bottom, GRAV, GINO, VEC and CHIRAL. The SU(3) Dynkin labels $[p,q]$ corresponding to each of these multiplets present in \eqref{eq:AllSU3Mults_ell0} and \eqref{eq:AllSU3Mults}, and its multiplicity $h$, can be read off from the term $h\, x^p\, y^q$ in the polynomial expansion of the corresponding fugacity. The concrete dictionary is
\begin{equation}
	x^a\, y^b\, \frac{y^{\ell+1}-x^{\ell+1}}{y-x} \; \longrightarrow \; 
	\bigoplus_{p=0}^{\ell}[p+a,\ell-p+b]\,.
\end{equation}

It is also useful to explicitly flesh out some of the analytic results for the spectrum given in section \ref{sec:U(1)2toSU3KK} for the family of solutions of \cite{Guarino:2019oct,Guarino:2020gfe} reviewed in section \ref{sec:U(1)2toSU3}. Let us do that, for example, for the KK tower with $\ell =0$ and all $n$, across the entire two-parameter family of solutions. In order to do this, we introduce the following notation
\begin{equation}	\label{eq: notationU1U1inSU3}
	{\rm MULT}[E_0;\, m_1,\, m_2\, (\bm{r})] \; ,
\end{equation}
for an OSp$(4|1)$ multiplet MULT of dimension $E_0$ given by (\ref{eq:GenericE0}), present in the spectrum at those KK levels, with $ \textrm{U}(1)_1 \times \textrm{U}(1)_2$ charges $m_1$, $m_2$ that derive from each representation $\bm{r}$ of SU(3) with Dynkin labels $[p,q]$ in (\ref{eq:AllSU3Mults_ell0}). With these conventions, the KK tower with $\ell =0$ and all $n$, for all values of $\chi_1$, $\chi_2$, is
\small
\begin{equation}\label{eq:(ell=0,n)}
                 \begin{aligned}
&\text{\underline{GRAV}}\Big[1+\sqrt{\tfrac{9}{4}+\tfrac{5}{4}(\tfrac{2\pi n}{T})^2};\, 0,\,0\, (\mathbf{1}) \Big]\,\\[10pt]
&\oplus\text{GINO}\Big[1+\zeta_{n,1,1};\, 1,\,1\, (\mathbf{3}) \Big]\oplus\text{GINO}\Big[1+\zeta_{n,-1,1};\, -1,\,1\, (\mathbf{3}) \Big]\oplus\text{GINO}\Big[1+\zeta_{n,0,-2};\, 0,\,-2\, (\mathbf{3}) \Big]\\
&\oplus\text{GINO}\Big[1+\zeta_{n,-1,-1};\, -1,\,-1\, (\bar{\mathbf{3}}) \Big]\oplus\text{GINO}\Big[1+\zeta_{n,1,-1};\, 1,\,-1\, (\bar{\mathbf{3}}) \Big]\oplus\text{GINO}\Big[1+\zeta_{n,0,2};\, 0,\,2\, (\bar{\mathbf{3}}) \Big]\\[10pt]
 &\oplus\text{VEC}\Big[1+\delta_{n,1,1};\, 1,\,1\, (\mathbf{3}) \Big]\oplus\text{VEC}\Big[1+\delta_{n,-1,1};\, -1,\,1\, (\mathbf{3}) \Big]\oplus\text{VEC}\Big[1+\delta_{n,0,-2};\, 0,\,-2\, (\mathbf{3}) \Big]\\
 &\oplus\text{VEC}\Big[1+\delta_{n,-1,-1};\, -1,\,-1\, (\bar{\mathbf{3}}) \Big]\oplus\text{VEC}\Big[1+\delta_{n,1,-1};\, 1,\,-1\, (\bar{\mathbf{3}}) \Big]\oplus\text{VEC}\Big[1+\delta_{n,0,2};\, 0,\,2\, (\bar{\mathbf{3}}) \Big]\\
&\oplus2\times\text{\underline{VEC}}\Big[1+\rho_{n,0,0};\, 0,\,0\, (\mathbf{8}) \Big]\\
&\quad\oplus\text{\underline{\underline{VEC}}}\Big[1+\rho_{n,2,0};\, 2,\, 0\, (\mathbf{8})\Big]\oplus\text{\underline{\underline{VEC}}}\Big[1+\rho_{n,-2,0};\, -2,\, 0\, (\mathbf{8})\Big]\\
&\qquad\oplus\text{\underline{\underline{\underline{VEC}}}}\Big[1+\rho_{n,1,3};\, 1,\,3\, (\mathbf{8}) \Big]\oplus\text{\underline{\underline{\underline{VEC}}}}\Big[1+\rho_{n,1,-3};\, 1,\,-3\, (\mathbf{8}) \Big]	\\
&\quad\qquad\oplus\text{\underline{\underline{\underline{VEC}}}}\Big[1+\rho_{n,-1,-3};\, -1,\,-3\, (\mathbf{8}) \Big]\oplus\text{\underline{\underline{\underline{VEC}}}}\Big[1+\rho_{n,-1,3};\, -1,\,3\, (\mathbf{8}) \Big]\\[10pt]
&\oplus2\times\text{CHIRAL}\Big[1+\sqrt{6+\tfrac{5}{4}\left(\tfrac{2\pi n}{T}\right)^2};\, 0,\, 0\, (\mathbf{1})\Big]\\
&\oplus\text{CHIRAL}\Big[1+\xi_{n,0,2};\, 0,\, 2\, (\mathbf{6})\Big]\oplus\text{CHIRAL}\Big[1+\xi_{n,0,-4};\, 0,\, -4\, (\mathbf{6})\Big]\\
&\quad\oplus \text{CHIRAL}\Big[1+\xi_{n,1,-1};\, 1,\, -1\, (\mathbf{6})\Big]\oplus \text{CHIRAL}\Big[1+\xi_{n,-1,-1};\, -1,\, -1\, (\mathbf{6})\Big] \\
&\qquad\oplus\text{CHIRAL}\Big[1+\xi_{n,2,2};\, 2,\, 2\, (\mathbf{6})\Big]\oplus\text{CHIRAL}\Big[1+\xi_{n,-2,2};\, -2,\, 2\, (\mathbf{6})\Big]\\
&\oplus\text{CHIRAL}\Big[1+\xi_{n,0,-2};\, 0,\,-2\, (\overline{ \mathbf{6}})\Big]\oplus\text{CHIRAL}\Big[1+\xi_{n,0,4};\, 0,\,4\, (\overline{ \mathbf{6}})\Big]\\
&\quad\oplus\text{CHIRAL}\Big[1+\xi_{n,-1,1};\, -1,\, 1\, (\overline{ \mathbf{6}})\Big]\oplus\text{CHIRAL}\Big[1+\xi_{n,1,1};\, 1,\,1\, (\overline{ \mathbf{6}})\Big]\\
&\qquad\oplus\text{CHIRAL}\Big[1+\xi_{n,-2,-2};\,-2,\,-2\, (\overline{ \mathbf{6}})\Big]\oplus\text{CHIRAL}\Big[1+\xi_{n,2,-2};\,2,\,-2\, (\overline{ \mathbf{6}})\Big]\, .\\
\end{aligned} 
\end{equation}
\normalsize
Here we have introduced the following shortand notations,
\begin{align}
&\nonumber\zeta_{n,m_1,m_2} \equiv \sqrt{\tfrac{16}{9}+\tfrac{5}{4} f_{nm_1m_2} (\chi_1 , \chi_2) }\,,\qquad\enspace\,
\delta_{n,m_1,m_2} \equiv  \sqrt{\tfrac{109}{36}+\tfrac{5}{4} f_{nm_1m_2} (\chi_1 , \chi_2) }\, ,\\
&\rho_{n,m_1,m_2} \equiv  \sqrt{\tfrac{1}{4}+\tfrac{5}{4} f_{nm_1m_2}  (\chi_1 , \chi_2) }\,, \qquad \quad
\xi_{n,m_1,m_2} \equiv  \sqrt{\tfrac{4}{9}+\tfrac{5}{4} f_{nm_1m_2}  (\chi_1 , \chi_2) }\, . \hspace{0.2cm}
\end{align} 
with $f_{nm_1m_2} (\chi_1 , \chi_2)$ given in  (\ref{eq:chiComb}).

At $n=0$, the underlined GRAV multiplet in (\ref{eq:(ell=0,n)}) splits as in (\ref{eq: shortening_generalGRAV}) for all values of the moduli. The resulting MGRAV contains the massless graviton. Similarly, the underlined VEC multiplet branches at $n=0$ and all $\chi_1$, $\chi_2$ as in (\ref{eq: shortening_generalVEC}). The resulting MVEC and CHIRALs respectively contain the $ \textrm{U}(1)_1 \times \textrm{U}(1)_2$ gauge fields and the moduli $\chi_1$, $\chi_2$. At an $\textrm{SU}(2)\times \textrm{U}(1)$ or SU$(3)$ invariant location, the doubly and triply  underlined VEC multiplets also split through (\ref{eq: shortening_generalVEC}). This happens at either $n=0$ or at a `space-invading' level $n^\prime \neq 0$ given by (\ref{eq:KKreshuffle}), depending on the conditions discussed in section \ref{sec:U(1)2toSU3KK}. In particular, the $\ell =n =0$ scalar and vector spectrum reported for this family in (2.25), (2.26) of \cite{Guarino:2020gfe} is recovered from (\ref{eq:(ell=0,n)}).  The $\ell=0$ and all $n$ spectrum summarised in table~\ref{table: (l=0,n)multipletsSU(3)} of the main text for the $\chi_1 = \chi_2 = 0$, SU(3)-invariant point, is also recovered from (\ref{eq:(ell=0,n)}). 

As a final check, we can recover the spectrum of gravitons for the $\chi_1 = \chi_2 = 0$, SU(3)-invariant solution, which was computed for all $\ell$ and $n$ in \cite{Dimmitt:2019qla} using the methods of \cite{Bachas:2011xa}. In table 4 of \cite{Dimmitt:2019qla}, the graviton masses were given as
\begin{equation}	\label{eq:N1u1u1gravs}
	L^2M^2=\tfrac{5}{6} \ell(\ell+4)-\tfrac{5}{36} (\ell-2 p)^2+\tfrac{5\pi^2}{T^2} n^2 	\; ,
\end{equation}
with $n_{\rm here} = j_{\rm there}$. Now, (\ref{eq:AllSU3Mults_ell0}), (\ref{eq:AllSU3Mults}) of the main text show that $\ell+1$ GRAV multiplets arise in the spectrum at fixed $\ell$ and all $n$ in the $[p, \ell - p]$ representation of SU(3), with $p = 0 , 1  , \ldots , \ell$. By some straightforward manipulation of (\ref{eq: conformaldimensions}) evaluated at $\chi_1 = \chi_2 =0$ and with these charges, these multiplets can be checked to have dimension
\begin{equation}
E_0 = 1 + \sqrt{ \tfrac94 + \tfrac{5}{6} \ell(\ell+4)-\tfrac{5}{36} (\ell-2 p)^2+\tfrac{5\pi^2}{T^2} n^2 } \; .
\end{equation}
Thus, the individual graviton states in these GRAV multiplets have dimension $\Delta = E_0 + \frac12$ (see {\it e.g.}~table 1 of \cite{Cesaro:2020soq}) and, indeed, mass (\ref{eq:N1u1u1gravs}) via $M^2L^2 = \Delta (\Delta-3)$. Curiously, the quantum numbers of these GRAVs conspire to have their dimensions depend on their charges, $\ell -2p$, under the broken U$(1)_\tau$ in (\ref{eq:Sol1Branching}) through the branching 
\begin{equation} \label{eq:SO6intoSO6U1IIB}
	[0,\ell,0]\xrightarrow[]{\text{SU(3)} \times\text{U(1)}_\tau}\bigoplus_{p=0}^\ell \, [p,\ell-p]_{\ell-2p} \, ,
\end{equation}	
under the first inclusion in (\ref{eq:Sol1Branching}).

\section{Some explicit type IIB uplifts} \label{sec:UpliftSU3Fam}

The family of type IIB solutions (\ref{eq:AdS4ixS5xS1}) corresponding to the two-parameter family of $D=4$ vacua \cite{Guarino:2019oct,Guarino:2020gfe} reviewed in section \ref{sec:U(1)2toSU3} was constructed in \cite{Guarino:2021kyp}. Here, we recover the uplift of this family starting from their KK spectrum, by the reverse engineering process outlined in section \ref{sec:Discussion}. In order to do this, it is helpful to use coordinates $\alpha$, $\theta$, $\phi$ $\psi$, $\tau$ on $S^5$ adapted to the Hopf fibration so that the Fubini-Study metric on the complex projective plane and the one-form $\sigma$ along the Hopf fibre take on the explicit form
\begin{equation}	\label{eq: CP2incoordinates}
	\begin{aligned}
		\sigma&=\tfrac12\sin^2\!\alpha\, (d\psi+\cos\theta\, d\phi)\,,	\\[5pt]
		ds^2(\mathbb{CP}^2)&=d\alpha^2+ \tfrac14 \, \sin^2 \alpha  \Big[d\theta^2+\sin^2\!\theta\,d\phi^2+\cos^2\!\alpha\,(d\psi+\cos\theta\, d\phi)^2\Big]\, .
	\end{aligned}
\end{equation}
It is also helpful to introduce a periodically identified coordinate $\eta \sim \eta +T$ on the S-folded $S^1$. The coordinates on $S^1 \times S^5$ range as
\begin{equation}	\label{eq: coordsSU3sol}
	0\leq\eta<T\,,		\quad
	-\frac{\pi}2\leq\alpha\leq\frac{\pi}2\,,	\quad
	0\leq\theta\leq\pi\,,	\quad
	0\leq\phi<2\pi\,,	\quad
	0\leq\psi<2\pi\, ,	\quad
	0\leq\tau<2\pi\,.
\end{equation}
In particular, $\phi$, $\psi$ and $\tau$ are periodic with period $2\pi$. It is also useful to introduce the frame
\begin{equation}	\label{eq: sechsbein}
	\begin{alignedat}{3}
		&e^1=d\alpha\,,	\quad\qquad	&& e^2=\tfrac{1}2\sin\alpha\cos\alpha (\bm{v}_\psi+\cos\theta\,\bm{v}_\phi)\,,	\quad\qquad	&&e^3=\tfrac{1}2\sin\alpha\, d\theta\,, \\[5pt]
		&e^4=-\tfrac{1}2\sin\alpha\sin\theta\, \bm{v}_\phi\,,	\quad\qquad	&& e^5=\bm{v}_\tau+\sigma'\, ,	\quad\qquad	&& e^{6}=d\eta\,, 
	\end{alignedat}
\end{equation}
where
\begin{equation}	\label{eq:sigmaPrimed}
		\sigma' =\tfrac12\sin^2\!\alpha\, \big(\bm{v}_\psi+\cos\theta\, \bm{v}_\phi\big)\, ,
\end{equation}
and
\begin{equation}	\label{eq: deformedSU3v}
	\bm{v}_{\phi}=d\phi-(\chi_1-\chi_2)d\eta\,,	\quad
	\bm{v}_{\psi}=d\psi-3(\chi_1+\chi_2)d\eta\,,	\quad
	\bm{v}_{\tau}=d\tau+(\chi_1+\chi_2)d\eta\, ,
\end{equation}
following \eqref{eq: deformedSU3}. Finally, it is helpful to write the following one- and two-forms
\begin{equation}	\label{eq: defSEforms}
	\bm{\eta}'=e^5		\,,\qquad
	\bm{J}'=e^1\wedge e^2+e^3\wedge e^4			\,,\qquad
	\bm{\Omega}'=	e^{i(3\tau+\psi)}(e^1+i\, e^2)\wedge(e^3+i\, e^4)	\, .
\end{equation}
These define a Sasaki-Einstein structure with
\begin{equation}	\label{eq: defSErelations}
	\bm{J}'\wedge\bm{\Omega}'=0	\,,\qquad
	\bm{\Omega}'\wedge\bm{\bar\Omega}'=2\,\bm{J}'\wedge\bm{J}' \,,\qquad
	d\bm{\eta}'=2\bm{J}'\,,\qquad 
	d\bm{\Omega}'=3i\,\bm{\eta}'\wedge\bm{\Omega}' \; .
\end{equation}

The type IIB uplift of the two-parameter family of solutions arises by applying the replacements (\ref{eq: deformedSU3}) to the SU(3)-invariant type IIB solution of \cite{Guarino:2019oct}. With the above definitions and in the type IIB conventions of appendix A of \cite{Gauntlett:2010vu}, the result is:
\begin{equation}	\label{eq:IIBSol}
	\begin{aligned}
	ds_{10}^2 &= \tfrac{3\sqrt6}5\,L^2 \, ds^2({\rm AdS}_4)+ \sqrt{\tfrac65} \,\Big[\tfrac56\,ds^2(\mathbb{CP}'^2)+(\bm{v}_\tau+\sigma')^2+\tfrac59\,d\eta^2\Big] \,, \\[5pt]
		\tilde{F}_5&=\tfrac{3}{2}(1+*')\,\bm{\eta}'\wedge\bm{J}'\wedge\bm{J}'\,,	\\[5pt]
		H_3+i\, F_3&=\tfrac1{\sqrt6}\Big(e^{6}\wedge\,\bm{\Omega}'+3\,e^5\wedge\,\bar{\bm{\Omega}}'\Big)\, ,\\[5pt]
		C_0+i\, e^{-\Phi} &= \frac{2\sinh\eta\,\cosh\eta+i}{\sinh^2\eta+\cosh^2\eta}\, ,
	\end{aligned}
\end{equation}
where $ds^2(\mathbb{CP}'^2)$ is the deformation of (\ref{eq: CP2incoordinates}) given by 
\begin{equation}	\label{eq: defCP2incoordinates}
		ds^2(\mathbb{CP}'^2)=d\alpha^2+ \tfrac14 \, \sin^2 \alpha  \Big[d\theta^2+\sin^2\!\theta\,\bm{v}_\phi^2+\cos^2\!\alpha\,(\bm{v}_\psi+\cos\theta\, \bm{v}_\phi)^2\Big]\,.
\end{equation}

The only fields in (\ref{eq:IIBSol}) that are independent of the parameters $\chi_1$, $\chi_2$ are the axion and the dilaton, $C_0$ and $\Phi$. At $\chi_1 = \chi_2 =0$, the configuration (\ref{eq:IIBSol}) reduces to the SU(3)-invariant solution given in \cite{Guarino:2019oct} after fixing some gauge redundancies there.  Locally, the dependence of (\ref{eq:IIBSol}) on the moduli $\chi_1$, $\chi_2$ can be removed by the change of coordinates
\begin{equation}	\label{eq: SU3primedcoords}
		\phi'=\phi-(\chi_1-\chi_2)\,\eta\,,	\qquad
		\psi'=\psi-3(\chi_1+\chi_2)\,\eta\,,	\qquad
		\tau'=\tau+(\chi_1+\chi_2)\,\eta\,.
\end{equation}
Thus, the solution (\ref{eq:IIBSol}) is locally equivalent to the SU(3)-invariant solution of \cite{Guarino:2019oct}. However, for generic values of the parameters, the coordinate transformation (\ref{eq: SU3primedcoords}) does not generate a diffeomorphism because it is not globally defined (as the primed coordinates will not typically have period $2\pi$ like the unprimed ones). Only when
\begin{equation}	\label{eq: periodspsi12v2}
	\chi_{1}=\tfrac{2\pi}{T}h_1\,,			\qquad
	\chi_{2}=\tfrac{2\pi}{T}h_2\,,			\qquad\qquad
	h_{1,2}\in\mathbb{Z}\, ,
\end{equation}
does (\ref{eq: SU3primedcoords}) correspond to a diffeomorphism, thus leading to a solution that is globally equivalent to the $\chi_1 = \chi_2 = 0$ SU(3)-invariant solution. This leads to the periodic identification of the moduli advertised in (\ref{eq:periodicity}) and seen in the KK spectrum of section (\ref{sec:U(1)2toSU3KK}). 

A consequence of the local equivalence of (\ref{eq:IIBSol}) and the SU(3)-invariant type IIB solution of \cite{Guarino:2019oct}  is that the former automatically satisfies the equations of motion, reported to hold for the latter in that reference. Generically, the solution (\ref{eq:IIBSol}) is only invariant under the U$(1)^2$ generated by the Killing vectors $k_3$ and $k_8$ specified below (\ref{eq:DerRep}). When $\chi_1 = \chi_2$ modulo (\ref{eq:periodicity}), $\bm{v}_\phi=d\phi$ from (\ref{eq: deformedSU3}) and the solution becomes invariant under the SU(2) that rotates the $S^2$ with metric  $d\theta^2+\sin^2\!\theta\,d\phi^2$ in (\ref{eq: defCP2incoordinates}), in addition to still being invariant under the U(1) generated by $k_8$. Finally, when $\chi_1 = \chi_2 =0$ modulo (\ref{eq:periodicity}), the solution is invariant under the SU(3) that rotates the Fubini-Study metric (\ref{eq: CP2incoordinates}). In all cases, the U$(1)_\tau$ in (\ref{eq:Sol1Branching}) generated by $\partial_\tau$ is an isometry of the ten-dimensional metric, but is broken in the full solution by the $\bm{\Omega}'$ dependence of the fluxes.

By similar arguments, we can determine the type IIB uplift of the one-parameter families of \cite{Berman:2021ynm,Bobev:2021rtg}. Denoting by $\beta$ the $S^5$ angle associated to the preserved U(1), the type IIB uplifts of these families are obtained from that of the $\chi =0$ representative by applying the shift (\ref{eq:shiftbeta}). In particular, the uplift of the family reviewed in section \ref{sec:U(1)toSO3} can be obtained simply by making the replacement (\ref{eq:shiftbeta}) in the one-forms $\Psi^1$, $\Psi^2$ and $\Psi^3$ defined in equation (3.37) of \cite{Berman:2021ynm}. The resulting $\chi$-dependent family of solutions of type IIB reduces to the SO(3)-invariant solution at $\chi =0$ given in that reference. The family is also locally equivalent to the latter, but only globally equivalent at the $\chi$ locations that lead to the indentification (\ref{eq:periodicitySO3Fam}).

\bibliography{references}

\providecommand{\href}[2]{#2}\begingroup\raggedright\begin{thebibliography}{10}

\bibitem{Gallerati:2014xra}
A.~Gallerati, H.~Samtleben, and M.~Trigiante, {\it {The $ \mathcal{N}>2 $
  supersymmetric AdS vacua in maximal supergravity}},  {\em JHEP} {\bf 12}
  (2014) 174, [\href{http://arxiv.org/abs/1410.0711}{{\tt arXiv:1410.0711}}].

\bibitem{Guarino:2019oct}
A.~Guarino and C.~Sterckx, {\it {S-folds and (non-)supersymmetric Janus
  solutions}},  {\em JHEP} {\bf 12} (2019) 113,
  [\href{http://arxiv.org/abs/1907.04177}{{\tt arXiv:1907.04177}}].

\bibitem{Guarino:2020gfe}
A.~Guarino, C.~Sterckx, and M.~Trigiante, {\it {$\mathcal{N}=2$ supersymmetric
  S-folds}},  {\em JHEP} {\bf 04} (2020) 050,
  [\href{http://arxiv.org/abs/2002.03692}{{\tt arXiv:2002.03692}}].

\bibitem{Bobev:2021yya}
N.~Bobev, F.~F. Gautason, and J.~van Muiden, {\it {The holographic conformal
  manifold of 3d $ \mathcal{N} $ = 2 S-fold SCFTs}},  {\em JHEP} {\bf 07}
  (2021), no.~221 221, [\href{http://arxiv.org/abs/2104.00977}{{\tt
  arXiv:2104.00977}}].

\bibitem{Guarino:2021hrc}
A.~Guarino and C.~Sterckx, {\it {Flat deformations of type IIB S-folds}},  {\em
  JHEP} {\bf 11} (2021) 171, [\href{http://arxiv.org/abs/2109.06032}{{\tt
  arXiv:2109.06032}}].

\bibitem{Berman:2021ynm}
D.~Berman, T.~Fischbacher, and G.~Inverso, {\it {New $ \mathcal{N} $ = 1
  AdS$_{4}$ solutions of type IIB supergravity}},  {\em JHEP} {\bf 03} (2022)
  097, [\href{http://arxiv.org/abs/2111.03002}{{\tt arXiv:2111.03002}}].

\bibitem{Bobev:2021rtg}
N.~Bobev, F.~F. Gautason, and J.~van Muiden, {\it {Holographic 3d $\mathcal
  N=1$ Conformal Manifolds}},  \href{http://arxiv.org/abs/2111.11461}{{\tt
  arXiv:2111.11461}}.

\bibitem{Inverso:2016eet}
G.~Inverso, H.~Samtleben, and M.~Trigiante, {\it {Type II supergravity origin
  of dyonic gaugings}},  {\em Phys. Rev.} {\bf D95} (2017), no.~6 066020,
  [\href{http://arxiv.org/abs/1612.05123}{{\tt arXiv:1612.05123}}].

\bibitem{DHoker:2007zhm}
E.~D'Hoker, J.~Estes, and M.~Gutperle, {\it {Exact half-BPS Type IIB interface
  solutions. I. Local solution and supersymmetric Janus}},  {\em JHEP} {\bf 06}
  (2007) 021, [\href{http://arxiv.org/abs/0705.0022}{{\tt arXiv:0705.0022}}].

\bibitem{Gaiotto:2008sd}
D.~Gaiotto and E.~Witten, {\it {Janus Configurations, Chern-Simons Couplings,
  And The theta-Angle in N=4 Super Yang-Mills Theory}},  {\em JHEP} {\bf 06}
  (2010) 097, [\href{http://arxiv.org/abs/0804.2907}{{\tt arXiv:0804.2907}}].

\bibitem{Bobev:2019jbi}
N.~Bobev, F.~F. Gautason, K.~Pilch, M.~Suh, and J.~Van~Muiden, {\it {Janus and
  J-fold Solutions from Sasaki-Einstein Manifolds}},  {\em Phys. Rev.} {\bf
  D100} (2019), no.~8 081901, [\href{http://arxiv.org/abs/1907.11132}{{\tt
  arXiv:1907.11132}}].

\bibitem{Bobev:2020fon}
N.~Bobev, F.~F. Gautason, K.~Pilch, M.~Suh, and J.~van Muiden, {\it
  {Holographic interfaces in $ \mathcal{N} $ = 4 SYM: Janus and J-folds}},
  {\em JHEP} {\bf 05} (2020) 134, [\href{http://arxiv.org/abs/2003.09154}{{\tt
  arXiv:2003.09154}}].

\bibitem{Arav:2021tpk}
I.~Arav, K.~C.~M. Cheung, J.~P. Gauntlett, M.~M. Roberts, and C.~Rosen, {\it {A
  new family of $AdS_4$ S-folds in type IIB string theory}},  {\em JHEP} {\bf
  05} (2021) 222, [\href{http://arxiv.org/abs/2101.07264}{{\tt
  arXiv:2101.07264}}].

\bibitem{Arav:2021gra}
I.~Arav, J.~P. Gauntlett, M.~M. Roberts, and C.~Rosen, {\it {Marginal
  deformations and RG flows for type IIB S-folds}},  {\em JHEP} {\bf 07} (2021)
  151, [\href{http://arxiv.org/abs/2103.15201}{{\tt arXiv:2103.15201}}].

\bibitem{deWit:1982ig}
B.~de~Wit and H.~Nicolai, {\it {N=8 Supergravity}},  {\em Nucl.Phys.} {\bf
  B208} (1982) 323.

\bibitem{Guarino:2015qaa}
A.~Guarino and O.~Varela, {\it {Dyonic ISO(7) supergravity and the duality
  hierarchy}},  {\em JHEP} {\bf 02} (2016) 079,
  [\href{http://arxiv.org/abs/1508.04432}{{\tt arXiv:1508.04432}}].

\bibitem{deWit:1986iy}
B.~de~Wit and H.~Nicolai, {\it {The Consistency of the $S^7$ Truncation in
  $D=11$ Supergravity}},  {\em Nucl.Phys.} {\bf B281} (1987) 211.

\bibitem{Guarino:2015jca}
A.~Guarino, D.~L. Jafferis, and O.~Varela, {\it {The string origin of dyonic
  N=8 supergravity and its simple Chern-Simons duals}},  {\em Phys. Rev. Lett.}
  {\bf 115} (2015), no.~9 091601, [\href{http://arxiv.org/abs/1504.08009}{{\tt
  arXiv:1504.08009}}].

\bibitem{Guarino:2015vca}
A.~Guarino and O.~Varela, {\it {Consistent $ \mathcal{N}=8 $ truncation of
  massive IIA on S$^{6}$}},  {\em JHEP} {\bf 12} (2015) 020,
  [\href{http://arxiv.org/abs/1509.02526}{{\tt arXiv:1509.02526}}].

\bibitem{Lunin:2005jy}
O.~Lunin and J.~M. Maldacena, {\it {Deforming field theories with U(1) x U(1)
  global symmetry and their gravity duals}},  {\em JHEP} {\bf 05} (2005) 033,
  [\href{http://arxiv.org/abs/hep-th/0502086}{{\tt hep-th/0502086}}].

\bibitem{Ashmore:2016oug}
A.~Ashmore, M.~Gabella, M.~Gra\~na, M.~Petrini, and D.~Waldram, {\it {Exactly
  marginal deformations from exceptional generalised geometry}},  {\em JHEP}
  {\bf 01} (2017) 124, [\href{http://arxiv.org/abs/1605.05730}{{\tt
  arXiv:1605.05730}}].

\bibitem{Bobev:2021gza}
N.~Bobev, P.~Bomans, F.~F. Gautason, and V.~S. Min, {\it {Marginal deformations
  from type IIA supergravity}},  {\em SciPost Phys.} {\bf 10} (2021), no.~6
  140, [\href{http://arxiv.org/abs/2103.02038}{{\tt arXiv:2103.02038}}].

\bibitem{Ashmore:2021mao}
A.~Ashmore, M.~Petrini, E.~L. Tasker, and D.~Waldram, {\it {Exactly Marginal
  Deformations and Their Supergravity Duals}},  {\em Phys. Rev. Lett.} {\bf
  128} (2022), no.~19 191601, [\href{http://arxiv.org/abs/2112.08375}{{\tt
  arXiv:2112.08375}}].

\bibitem{Assel:2018vtq}
B.~Assel and A.~Tomasiello, {\it {Holographic duals of 3d S-fold CFTs}},  {\em
  JHEP} {\bf 06} (2018) 019, [\href{http://arxiv.org/abs/1804.06419}{{\tt
  arXiv:1804.06419}}].

\bibitem{Giambrone:2021zvp}
A.~Giambrone, E.~Malek, H.~Samtleben, and M.~Trigiante, {\it {Global properties
  of the conformal manifold for S-fold backgrounds}},  {\em JHEP} {\bf 06}
  (2021), no.~111 111, [\href{http://arxiv.org/abs/2103.10797}{{\tt
  arXiv:2103.10797}}].

\bibitem{Cesaro:2021tna}
M.~Ces\`aro, G.~Larios, and O.~Varela, {\it {The spectrum of
  marginally-deformed $ \mathcal{N} $ = 2 CFTs with AdS$_{4}$ S-fold duals of
  type IIB}},  {\em JHEP} {\bf 12} (2021) 214,
  [\href{http://arxiv.org/abs/2109.11608}{{\tt arXiv:2109.11608}}].

\bibitem{Guarino:2021kyp}
A.~Guarino and C.~Sterckx, {\it {S-folds and holographic RG flows on the
  D3-brane}},  {\em JHEP} {\bf 06} (2021) 051,
  [\href{http://arxiv.org/abs/2103.12652}{{\tt arXiv:2103.12652}}].

\bibitem{Malek:2019eaz}
E.~Malek and H.~Samtleben, {\it {Kaluza-Klein Spectrometry for Supergravity}},
  {\em Phys. Rev. Lett.} {\bf 124} (2020), no.~10 101601,
  [\href{http://arxiv.org/abs/1911.12640}{{\tt arXiv:1911.12640}}].

\bibitem{Malek:2020yue}
E.~Malek and H.~Samtleben, {\it {Kaluza-Klein Spectrometry from Exceptional
  Field Theory}},  {\em Phys. Rev. D} {\bf 102} (2020), no.~10 10,
  [\href{http://arxiv.org/abs/2009.03347}{{\tt arXiv:2009.03347}}].

\bibitem{Varela:2020wty}
O.~Varela, {\it {Super-Chern-Simons spectra from Exceptional Field Theory}},
  {\em JHEP} {\bf 04} (2021) 283, [\href{http://arxiv.org/abs/2010.09743}{{\tt
  arXiv:2010.09743}}].

\bibitem{Cesaro:2020soq}
M.~Cesaro and O.~Varela, {\it {Kaluza-Klein fermion mass matrices from
  exceptional field theory and $ \mathcal{N} $ = 1 spectra}},  {\em JHEP} {\bf
  03} (2021) 138, [\href{http://arxiv.org/abs/2012.05249}{{\tt
  arXiv:2012.05249}}].

\bibitem{Hohm:2013pua}
O.~Hohm and H.~Samtleben, {\it {Exceptional Form of D=11 Supergravity}},  {\em
  Phys.Rev.Lett.} {\bf 111} (2013) 231601,
  [\href{http://arxiv.org/abs/1308.1673}{{\tt arXiv:1308.1673}}].

\bibitem{Hohm:2013uia}
O.~Hohm and H.~Samtleben, {\it {Exceptional Field Theory II: E$_{7(7)}$}},
  {\em Phys.Rev.} {\bf D89} (2014), no.~6 066017,
  [\href{http://arxiv.org/abs/1312.4542}{{\tt arXiv:1312.4542}}].

\bibitem{Godazgar:2014nqa}
H.~Godazgar, M.~Godazgar, O.~Hohm, H.~Nicolai, and H.~Samtleben, {\it
  {Supersymmetric E$_{7(7)}$ Exceptional Field Theory}},  {\em JHEP} {\bf 09}
  (2014) 044, [\href{http://arxiv.org/abs/1406.3235}{{\tt arXiv:1406.3235}}].

\bibitem{Berman:2020tqn}
D.~S. Berman and C.~D.~A. Blair, {\it {The Geometry, Branes and Applications of
  Exceptional Field Theory}},  {\em Int. J. Mod. Phys. A} {\bf 35} (2020),
  no.~30 2030014, [\href{http://arxiv.org/abs/2006.09777}{{\tt
  arXiv:2006.09777}}].

\bibitem{Dimmitt:2019qla}
K.~Dimmitt, G.~Larios, P.~Ntokos, and O.~Varela, {\it {Universal properties of
  Kaluza-Klein gravitons}},  {\em JHEP} {\bf 03} (2020) 039,
  [\href{http://arxiv.org/abs/1911.12202}{{\tt arXiv:1911.12202}}].

\bibitem{Englert:1983rn}
F.~Englert and H.~Nicolai, {\it {Supergravity in eleven-dimensional
  space-time}},  in {\em {12th International Colloquium on Group Theoretical
  Methods in Physics}}, pp.~249--283, 9, 1983.

\bibitem{Sezgin:1983ik}
E.~Sezgin, {\it {The Spectrum of the Eleven-dimensional Supergravity
  Compactified on the Round Seven Sphere}},  {\em Phys. Lett. B} {\bf 138}
  (1984) 57--62.

\bibitem{Biran:1983iy}
B.~Biran, A.~Casher, F.~Englert, M.~Rooman, and P.~Spindel, {\it {The
  Fluctuating Seven Sphere in Eleven-dimensional Supergravity}},  {\em Phys.
  Lett. B} {\bf 134} (1984) 179.

\bibitem{Klebanov:2008vq}
I.~Klebanov, T.~Klose, and A.~Murugan, {\it {AdS(4)/CFT(3) Squashed, Stretched
  and Warped}},  {\em JHEP} {\bf 03} (2009) 140,
  [\href{http://arxiv.org/abs/0809.3773}{{\tt arXiv:0809.3773}}].

\bibitem{Cesaro:2021haf}
M.~Cesaro, G.~Larios, and O.~Varela, {\it {Supersymmetric spectroscopy on
  $\textrm{AdS}_4 \times S^7$ and $\textrm{AdS}_4 \times S^6$}},  {\em JHEP}
  {\bf 07} (2021) 094, [\href{http://arxiv.org/abs/2103.13408}{{\tt
  arXiv:2103.13408}}].

\bibitem{Malek:2020mlk}
E.~Malek, H.~Nicolai, and H.~Samtleben, {\it {Tachyonic Kaluza-Klein modes and
  the AdS swampland conjecture}},  {\em JHEP} {\bf 08} (2020) 159,
  [\href{http://arxiv.org/abs/2005.07713}{{\tt arXiv:2005.07713}}].

\bibitem{Guarino:2020flh}
A.~Guarino, E.~Malek, and H.~Samtleben, {\it {Stable Nonsupersymmetric
  Anti\textendash{}de Sitter Vacua of Massive IIA Supergravity}},  {\em Phys.
  Rev. Lett.} {\bf 126} (2021), no.~6 061601,
  [\href{http://arxiv.org/abs/2011.06600}{{\tt arXiv:2011.06600}}].

\bibitem{Eloy:2020uix}
C.~Eloy, {\it {Kaluza-Klein spectrometry for ${\rm AdS_{3}}$ vacua}},  {\em
  SciPost Phys.} {\bf 10} (2021), no.~6 131,
  [\href{http://arxiv.org/abs/2011.11658}{{\tt arXiv:2011.11658}}].

\bibitem{Bobev:2020lsk}
N.~Bobev, E.~Malek, B.~Robinson, H.~Samtleben, and J.~van Muiden, {\it
  {Kaluza-Klein Spectroscopy for the Leigh-Strassler SCFT}},  {\em JHEP} {\bf
  04} (2021) 208, [\href{http://arxiv.org/abs/2012.07089}{{\tt
  arXiv:2012.07089}}].

\bibitem{Eloy:2021fhc}
C.~Eloy, G.~Larios, and H.~Samtleben, {\it {Triality and the consistent
  reductions on $\textrm{AdS}_4 \times S^3$}},  {\em JHEP} {\bf 01} (2022) 055,
  [\href{http://arxiv.org/abs/2111.01167}{{\tt arXiv:2111.01167}}].

\bibitem{Giambrone:2021wsm}
A.~Giambrone, A.~Guarino, E.~Malek, H.~Samtleben, C.~Sterckx, and M.~Trigiante,
  {\it {Holographic evidence for nonsupersymmetric conformal manifolds}},  {\em
  Phys. Rev. D} {\bf 105} (2022), no.~6 066018,
  [\href{http://arxiv.org/abs/2112.11966}{{\tt arXiv:2112.11966}}].

\bibitem{DallAgata:2011aa}
G.~Dall'Agata and G.~Inverso, {\it {On the Vacua of N = 8 Gauged Supergravity
  in 4 Dimensions}},  {\em Nucl.Phys.} {\bf B859} (2012) 70--95,
  [\href{http://arxiv.org/abs/1112.3345}{{\tt arXiv:1112.3345}}].

\bibitem{Dall'Agata:2014ita}
G.~Dall'Agata, G.~Inverso, and A.~Marrani, {\it {Symplectic Deformations of
  Gauged Maximal Supergravity}},  {\em JHEP} {\bf 1407} (2014) 133,
  [\href{http://arxiv.org/abs/1405.2437}{{\tt arXiv:1405.2437}}].

\bibitem{Inverso:2015viq}
G.~Inverso, {\it {Electric-magnetic deformations of D = 4 gauged
  supergravities}},  {\em JHEP} {\bf 03} (2016) 138,
  [\href{http://arxiv.org/abs/1512.04500}{{\tt arXiv:1512.04500}}].

\bibitem{deWit:2007mt}
B.~de~Wit, H.~Samtleben, and M.~Trigiante, {\it {The Maximal D=4
  supergravities}},  {\em JHEP} {\bf 0706} (2007) 049,
  [\href{http://arxiv.org/abs/0705.2101}{{\tt arXiv:0705.2101}}].

\bibitem{Trigiante:2016mnt}
M.~Trigiante, {\it {Gauged Supergravities}},  {\em Phys. Rept.} {\bf 680}
  (2017) 1--175, [\href{http://arxiv.org/abs/1609.09745}{{\tt
  arXiv:1609.09745}}].

\bibitem{Duff:1986hr}
M.~Duff, B.~Nilsson, and C.~Pope, {\it {Kaluza-Klein Supergravity}},  {\em
  Phys. Rept.} {\bf 130} (1986) 1--142.

\bibitem{Cesaro:2020piw}
M.~Cesaro, G.~Larios, and O.~Varela, {\it {A Cubic Deformation of ABJM: The
  Squashed, Stretched, Warped, and Perturbed Gets Invaded}},  {\em JHEP} {\bf
  10} (2020) 041, [\href{http://arxiv.org/abs/2007.05172}{{\tt
  arXiv:2007.05172}}].

\bibitem{Bachas:2011xa}
C.~Bachas and J.~Estes, {\it {Spin-2 spectrum of defect theories}},  {\em JHEP}
  {\bf 06} (2011) 005, [\href{http://arxiv.org/abs/1103.2800}{{\tt
  arXiv:1103.2800}}].

\bibitem{Larios:2019kbw}
G.~Larios, P.~Ntokos, and O.~Varela, {\it {Embedding the SU(3) sector of SO(8)
  supergravity in $D=11$}},  {\em Phys.\ Rev.\ D} {\bf 100} (2019), no.~8
  086021, [\href{http://arxiv.org/abs/1907.02087}{{\tt arXiv:1907.02087}}].

\bibitem{Gauntlett:2010vu}
J.~P. Gauntlett and O.~Varela, {\it {Universal Kaluza-Klein reductions of type
  IIB to N=4 supergravity in five dimensions}},  {\em JHEP} {\bf 1006} (2010)
  081, [\href{http://arxiv.org/abs/1003.5642}{{\tt arXiv:1003.5642}}].

\end{thebibliography}\endgroup

\end{document}